\documentclass[prd,twocolumn,showpacs,amsmath,amssymb,tightenlines]{revtex4-2}
\usepackage{overpic}
\usepackage{graphicx}
\usepackage{epsfig}
\usepackage{dcolumn}
\usepackage{bm}
\usepackage{rotating}
\usepackage{multirow}
\usepackage{subfigure}
\usepackage{hhline}
\usepackage{amssymb}
\usepackage{lineno}
\usepackage[colorlinks,linkcolor=blue,anchorcolor=black,citecolor=blue]{hyperref}

\newcommand{\EE}{e^+e^-}

\graphicspath{{./Img/}}

\begin{document}
	\normalsize
	\parskip=5pt plus 1pt minus 1pt
	\hyphenpenalty=10000
	\tolerance=1000

\title{\boldmath Search for the production of deuterons and antideuterons in $e^+e^-$ annihilation at center-of-mass energies between 4.13 and 4.70~GeV}

\author{
\begin{small}
\begin{center}
M.~Ablikim$^{1}$, M.~N.~Achasov$^{13,b}$, P.~Adlarson$^{73}$, R.~Aliberti$^{34}$, A.~Amoroso$^{72A,72C}$, M.~R.~An$^{38}$, Q.~An$^{69,56}$, Y.~Bai$^{55}$, O.~Bakina$^{35}$, I.~Balossino$^{29A}$
% ...
\\
\vspace{0.2cm}
(BESIII Collaboration)\\
\vspace{0.2cm} {\it
$^{1}$ Institute of High Energy Physics, Beijing 100049, People's Republic of China\\
% ...
$^{13}$ G.I. Budker Institute of Nuclear Physics SB RAS (BINP), Novosibirsk 630090, Russia\\
% ...
$^{29}$ INFN Sezione di Ferrara, (A)INFN Sezione di Ferrara, I-44122, Ferrara, Italy; (B)University of Ferrara,  I-44122, Ferrara, Italy\\
% ...
$^{34}$ Johannes Gutenberg University of Mainz, Johann-Joachim-Becher-Weg 45, D-55099 Mainz, Germany\\
$^{35}$ Joint Institute for Nuclear Research, 141980 Dubna, Moscow region, Russia\\
% ...
$^{38}$ Liaoning Normal University, Dalian 116029, People's Republic of China\\
% ...
$^{55}$ Southeast University, Nanjing 211100, People's Republic of China\\
$^{56}$ State Key Laboratory of Particle Detection and Electronics, Beijing 100049, Hefei 230026, People's Republic of China\\
% ...
$^{69}$ University of Science and Technology of China, Hefei 230026, People's Republic of China\\
% ...
$^{72}$ University of Turin and INFN, (A)University of Turin, I-10125, Turin, Italy; (B)University of Eastern Piedmont, I-15121, Alessandria, Italy; (C)INFN, I-10125, Turin, Italy\\
$^{73}$ Uppsala University, Box 516, SE-75120 Uppsala, Sweden\\
% ...
$^{b}$ Also at the Novosibirsk State University, Novosibirsk, 630090, Russia\\
% ...
}
\end{center}
\vspace{0.4cm}
\end{small}
 }
\noaffiliation{}

\date{\today}

\begin{abstract}

Using a data sample of $e^+e^-$ collision data corresponding to an
integrated luminosity of 19~fb$^{-1}$ collected with the BESIII
detector at the BEPCII collider, we search for the production of
deuterons and antideuterons via $e^+e^-\to pp\pi^-\bar{d}+c.c.$ for
the first time at center-of-mass energies between 4.13 and
4.70~GeV. No significant signal is observed and the upper limit of the
$e^+e^-\to pp\pi^-\bar{d}+c.c.$ cross section is determined to be from
9.0 to 145~fb depending on the center-of-mass energy at the $90\%$
confidence level.

\end{abstract}

\maketitle

\section{Introduction}
In the conventional quark model, mesons are composed of one quark and
one antiquark, while baryons are composed of three quarks.
However, many states with properties inconsistent with the conventional two or three quark models, such as the X(3872)~\cite{Belle:2003nnu},
Y(4260)~\cite{BaBar:2005hhc}, $Z_c$(3900)~\cite{BESIII:2013ris,Belle:2013yex},
$\pi_1(1600)$~\cite{E852:1998mbq}, and $P_c(4450)$~\cite{LHCb:2015yax}, have been discovered in the last two decades. Numerous theoretical proposals and experimental investigations on the subject of exotic states are reviewed in Refs.~\cite{Guo:2017jvc,Liu:2019zoy,Brambilla:2019esw,Klempt:2007cp,Jaffe:2004ph}, and there is strong evidence for the existence of tetraquark, pentaquark, and meson-meson and meson-baryon
molecular states~\cite{Guo:2017jvc,Liu:2019zoy,Brambilla:2019esw}.

The study of hadronic states with six quarks, either compact hexaquark
states or dibaryon states, has a long history~\cite{Locher:1985nu,Jaffe:2004ph,Abud:2009rk,Bashkanov:2013cla,Clement:2016vnl,Gal:2015rev}. Among them the $d^*(2380)$ has attracted substantial attention~\cite{Dong:2023xdi}.
The $d^*(2380)$, with a mass of about 2380~MeV and a width of about 70~MeV, was first observed in the isoscalar double-pionic fusion
process $pn\to d\pi^0\pi^0$~\cite{Bashkanov:2008ih},
and was later confirmed in other double-pionic fusion processes
$pn\to d\pi^+\pi^-$~\cite{WASA-at-COSY:2011bjg} and
$pp\to d\pi^+\pi^0$~\cite{CELSIUSWASA:2009dos},
and non-fusion processes
$pn\to pp\pi^0\pi^-$~\cite{WASA-at-COSY:2013fzt} and
$pn\to pn\pi^0\pi^0$~\cite{WASA-at-COSY:2014qkg}.
The $d^*(2380)$ has been proposed
to be the excited deuteron~($d$), a molecule with large
$\Delta\Delta$ component~\cite{Huang:2013nba}, or a hexaquark state which is dominated by the hidden-color component~\cite{Kim:2020rwn}.

The A2 collaboration observed a high spin polarization in the measurement of the recoiling neutron in deuterium photodisintegration, which could related to the excitation of the $d^*(2380)$~\cite{A2:2019arr}.  Apart from this measurement, most of the results related to the $d^*(2380)$ have so far come from the WASA experiment, and further studies from other experiments are needed to
confirm the existence of the $d^*(2380)$ and to search for
other similar states. The production of the antideuteron~($\bar{d}$) has been studied in several
$\EE$ annihilation experiments. The ARGUS~\cite{ARGUS:1989sto} and
CLEO~\cite{CLEO:2006zjy} experiments observed $\bar{d}$ production
at the level of $3\times 10^{-5}$ per $\Upsilon(1S)$ and $\Upsilon(2S)$
decays, and set limits on production in $\Upsilon(4S)$ decays and
$e^+e^-\to q\bar{q}$ at a center-of-mass energy ($\sqrt{s}$)
of 10.6~GeV. The BaBar~\cite{BaBar:2014ssg} experiment performed
measurements of inclusive antideuteron production in $\Upsilon(1S)$,
$\Upsilon(2S)$, $\Upsilon(3S)$ decays and in $e^+e^-$ annihilation
into hadrons at $\sqrt{s}\approx 10.58$~GeV.
The ALEPH~\cite{ALEPH:2006qoi} experiment observed
evidence at $3\sigma$ significance for $\bar{d}$ production in $e^+e^-\to q\bar{q}$ at
$\sqrt{s}=91.2$~GeV. Technically, the $d^*(2380)$ state could be studied
with these data by combining the detected antideuteron with pions
in the same event.

The BESIII experiment~\cite{BESIII:2009fln} collects $\EE$ collision
data at $\sqrt{s}$ between 2 and 4.95~GeV.  In recent years, the BESIII Collaboration has reported the observation
of $e^+e^- \to 2(p\bar{p})$~\cite{BESIII:2020svk} and $e^+e^- \to pp\bar{p}\bar{n}\pi^-+c.c.$~\cite{BESIII:pppnpi} with the production
cross sections on the order of 10~fb. This suggests the production of a deuteron or antideuteron together with two other nucleons may also be observed. 
In this paper, we search for the
production of (anti)deuterons at the BESIII experiment for the first time
using the process
$e^+e^-\to pp\pi^-\bar{d}+c.c.$ at $\sqrt{s}$ from 4.13 to 4.70~GeV.

\section{The BESIII detector and data samples}
The BESIII detector~\cite{BESIII:2009fln} records symmetric $e^+e^-$ collisions provided by the BEPCII storage ring~\cite{Yu:IPAC2016-TUYA01}, which operates in the $\sqrt{s}$ range from 2.0 to 4.95~GeV.
BESIII has collected large data samples in this energy region~\cite{BESIII:2020nme}. The cylindrical core of the BESIII detector covers 93\% of the full solid angle and consists of a helium-based multilayer drift chamber~(MDC), a plastic scintillator time-of-flight system~(TOF), and a CsI(Tl) electromagnetic calorimeter~(EMC), which are all enclosed in a superconducting solenoidal magnet providing a 1.0~T magnetic field~\cite{Huang:2022wuo}. The solenoid is supported by an octagonal flux-return yoke with resistive plate counter muon identification modules interleaved with steel. The charged-particle momentum resolution at $1~{\rm GeV}/c$ is $0.5\%$, and the specific energy loss (d$E$/d$x$) resolution is $6\%$ for electrons from Bhabha scattering. The EMC measures photon energies with a resolution of $2.5\%$ ($5\%$) at $1$~GeV in the barrel (end cap) region. The time resolution in the TOF barrel region is 68~ps, while that in the end cap region is 110~ps. The end cap TOF system was upgraded in 2015 using multi-gap resistive plate chamber technology, providing a time resolution of 60~ps~\cite{etof1,etof2,etof3}.

The data used in this work are listed in Table~\ref{Tab:Data_sample}.
The $\sqrt{s}$ for each data set was measured using the di-muon process~($e^+e^-\to\mu^+\mu^-$) with an uncertainty of less than 1.0 MeV~\cite{BESIII:2015zbz,BESIII:2020eyu,BESIII:2022ulv}, and the integrated luminosities were measured using the Bhabha process~($e^+e^-\to e^+e^-$) with an uncertainty of 1.0\%~\cite{BESIII:2022dxl,BESIII:2022ulv,BESIII:2015qfd}. The total integrated luminosity of the data used in this work is approximately 19~{\rm fb}$^{-1}$, of which about 15~{\rm fb}$^{-1}$ data samples were collected after the upgrade of the end cap TOF system.

\begin{table*}[htbp]
    \setlength{\abovecaptionskip}{0.cm}
    \setlength{\belowcaptionskip}{2pt}
    \centering
    \normalsize	
    \caption{The center-of-mass energies ($\sqrt{s}$), integrated luminosities ($\mathcal{L}_{\rm int}$), and signal yields for different cases. The uncertainties for signal yields are statistical only.}							
    \setlength{\tabcolsep}{12pt}			
    \renewcommand{\arraystretch}{1.2}		
	\begin{tabular}{cccccc}
		\hline
		 \multirow{2}{*}{$\sqrt{s}$ (MeV)}  & \multirow{2}{*}{$\mathcal{L}_{\rm int}$ (pb$^{-1}$)}   & \multicolumn{4}{c}{Signal yield}   \\
		\cline{3-6}
	      &  & case 1   & case 2   & case 3   & case 4   \\
    \hline
		4128.5        &401.5     &$\ \ \, 0.4\pm1.9$    &$0.0^{+0.9}$            &$0.0^{+0.9}$            &$0.0^{+0.9}$\\
		4157.4        &408.7     &$\ \ \, 1.0\pm2.6$    &$0.0^{+0.9}$            &$0.0^{+0.9}$            &$0.0^{+0.9}$\\
		4178.0       &3189.0     &$\ \ \ \; 5.7\pm11.7$    &$0.0^{+0.9}$            &$0.0^{+0.9}$            &$0.0^{+0.9}$\\
		4188.8        &526.7     &$\ \ \, 4.7\pm4.4$    &$0.0^{+0.9}$            &$0.0^{+0.9}$            &$0.0^{+0.9}$\\
		4198.9        &526.0     &$-0.1\pm3.6$    &$0.0^{+0.9}$            &$0.0^{+0.9}$            &$0.0^{+0.9}$\\
		4209.2        &517.1     &$-1.4\pm5.1$    &$0.0^{+0.9}$            &$1.0^{+1.4}$            &$0.0^{+0.9}$\\
		4218.7        &514.6     &$\ \, -2.1\pm12.6$   &$0.0^{+0.9}$            &$0.0^{+0.9}$            &$0.0^{+0.9}$\\
		4226.3       &1100.9     &$-0.7\pm4.6$    &$0.0^{+0.9}$            &$0.0^{+0.9}$            &$0.0^{+0.9}$\\
		4235.7        &530.3     &$\ \ \,3.4\pm4.4$    &$0.0^{+0.9}$            &$0.0^{+0.9}$            &$0.0^{+0.9}$\\
		4243.8        &538.1     &$-2.1\pm2.2$    &$0.0^{+0.9}$            &$0.0^{+0.9}$            &$0.0^{+0.9}$\\
		4258.0        &828.4     &$-2.1\pm4.5$    &$0.0^{+0.9}$            &$0.0^{+0.9}$            &$0.0^{+0.9}$\\
		4266.8        &531.1     &$\ \ \, 3.7\pm4.8$    &$0.0^{+0.9}$            &$0.0^{+0.9}$            &$0.0^{+0.9}$\\
		4277.7        &175.5     &$\ \ \, 0.0\pm3.0$     &$0.0^{+0.9}$            &$0.0^{+0.9}$            &$0.0^{+0.9}$\\
		4287.9        &502.4     &$-2.1\pm8.3$    &$0.0^{+0.9}$            &$0.0^{+0.9}$            &$0.0^{+0.9}$\\
		4312.0        &501.2     &$\ \ \, 0.9\pm3.4$    &$0.0^{+0.9}$            &$1.0^{+1.4}$            &$0.0^{+0.9}$\\
		4337.4        &505.0     &$\ \ \, 3.6\pm3.8$    &$0.0^{+0.9}$            &$0.0^{+0.9}$            &$0.0^{+0.9}$\\
		4358.3        &543.9     &$-2.0\pm7.1$    &$0.0^{+0.9}$            &$0.0^{+0.9}$            &$0.0^{+0.9}$\\
		4377.4        &522.7     &$\ \ \, 1.8\pm3.8$    &$0.0^{+0.9}$            &$0.0^{+0.9}$            &$0.0^{+0.9}$\\
		4396.4        &507.8     &$-1.8\pm6.3$    &$0.0^{+0.9}$            &$0.0^{+0.9}$            &$0.0^{+0.9}$\\
		4415.6       &1090.6     &$-1.0\pm3.8$    &$1.0_{-0.7}^{+1.4}$     &$1.0_{-0.7}^{+1.4}$     &$0.0^{+0.9}$\\
		4436.2        &569.9     &$\ \ \, 6.1\pm5.1$    &$0.0^{+0.9}$            &$0.0^{+0.9}$            &$0.0^{+0.9}$\\
		4467.1        &111.1     &$-2.1\pm0.6$    &$0.0^{+0.9}$            &$0.0^{+0.9}$            &$0.0^{+0.9}$\\
		4527.1        &112.1     &$-2.1\pm1.6$    &$0.0^{+0.9}$            &$0.0^{+0.9}$            &$0.0^{+0.9}$\\
		4599.5        &586.9     &$\ \ \, 1.4\pm4.3$    &$0.0^{+0.9}$            &$0.0^{+0.9}$            &$0.0^{+0.9}$\\
		4611.9	      &103.8     &$-2.1\pm1.0$    &$0.0^{+0.9}$            &$0.0^{+0.9}$            &$0.0^{+0.9}$\\
		4628.0	      &521.5     &$-2.1\pm1.5$    &$0.0^{+0.9}$            &$0.0^{+0.9}$            &$0.0^{+0.9}$\\
		4640.9        &552.4     &$-0.4\pm3.0$    &$1.0_{-0.7}^{+1.4}$     &$0.0^{+0.9}$            &$0.0^{+0.9}$\\
		4661.2        &529.6     &$-2.1\pm1.4$    &$0.0^{+0.9}$            &$0.0^{+0.9}$            &$0.0^{+0.9}$\\
		4681.9       &1669.3     &$-2.1\pm2.6$    &$0.0^{+0.9}$            &$0.0^{+0.9}$            &$0.0^{+0.9}$\\
		4698.8        &536.4     &$\ \ \, 3.3\pm3.5$    &$0.0^{+0.9}$            &$0.0^{+0.9}$            &$0.0^{+0.9}$\\
	\hline
	\end{tabular}
	\label{Tab:Data_sample}
\end{table*}

\begin{figure*}[htpb]
\centering
\includegraphics[angle=0,width=0.43\textwidth]{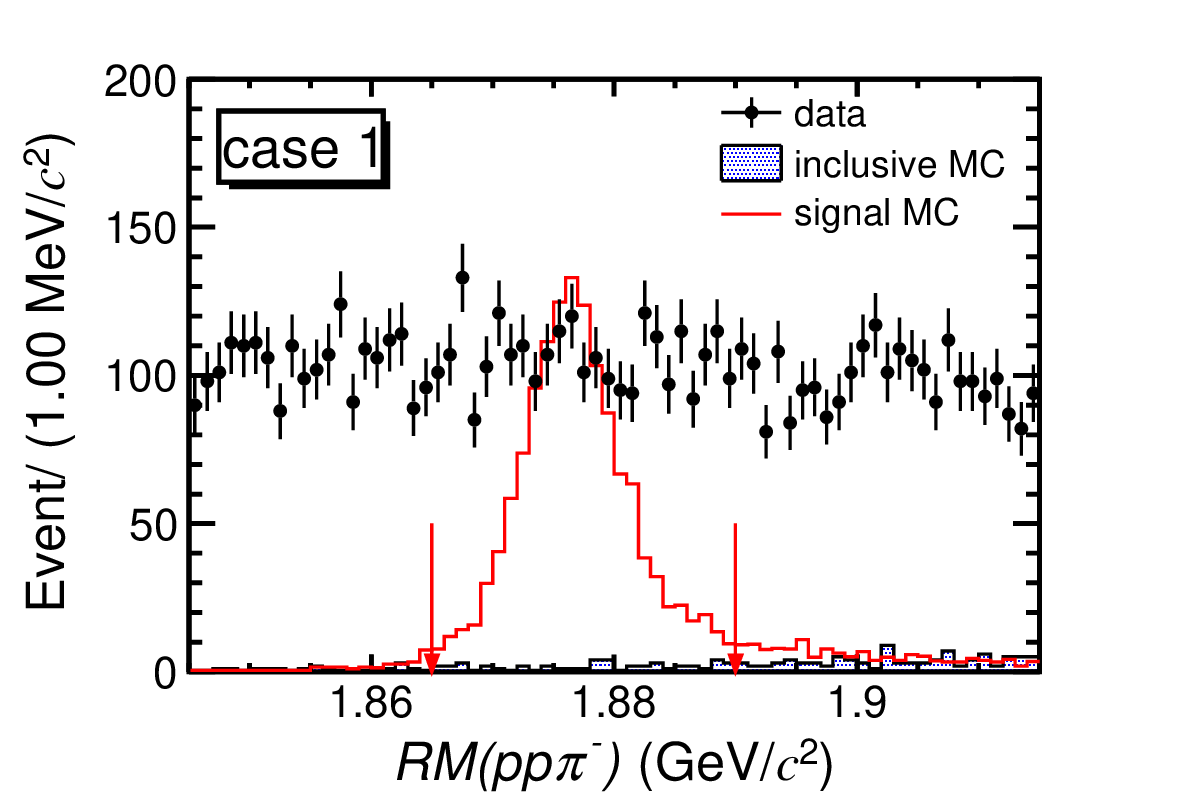}
\includegraphics[angle=0,width=0.43\textwidth]{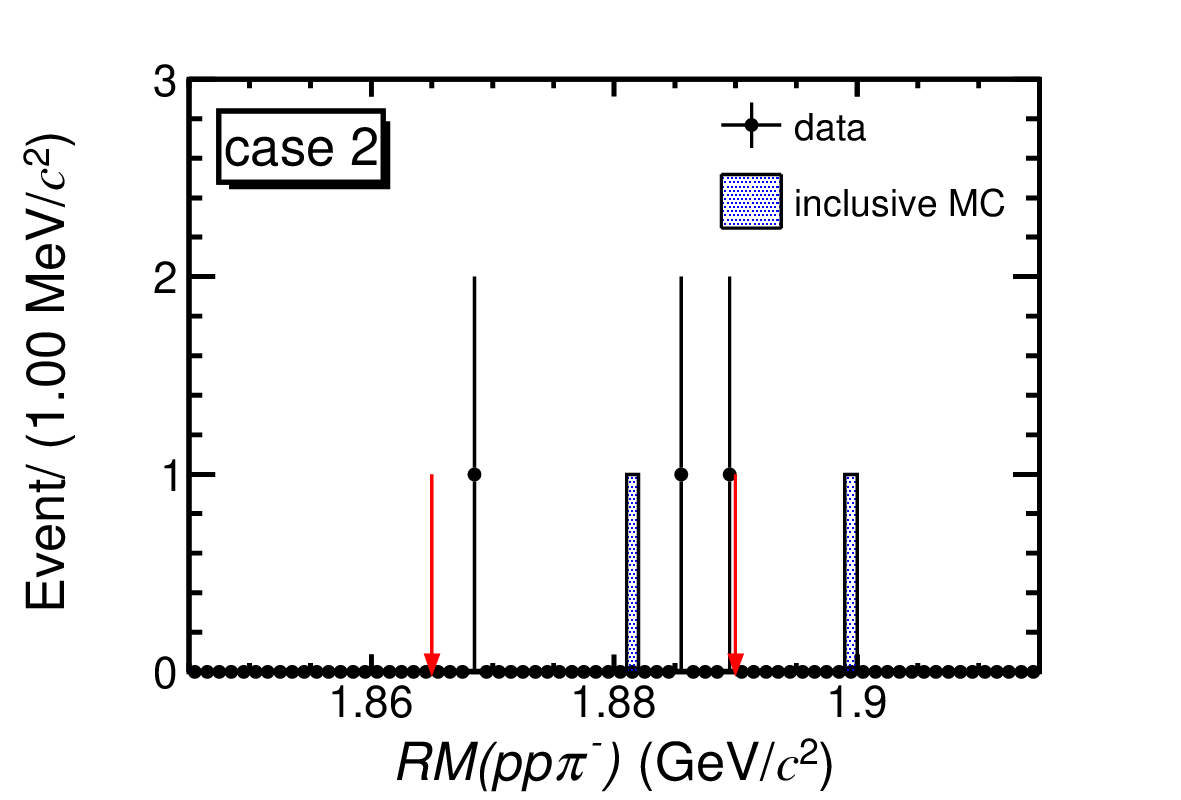}
\includegraphics[angle=0,width=0.43\textwidth]{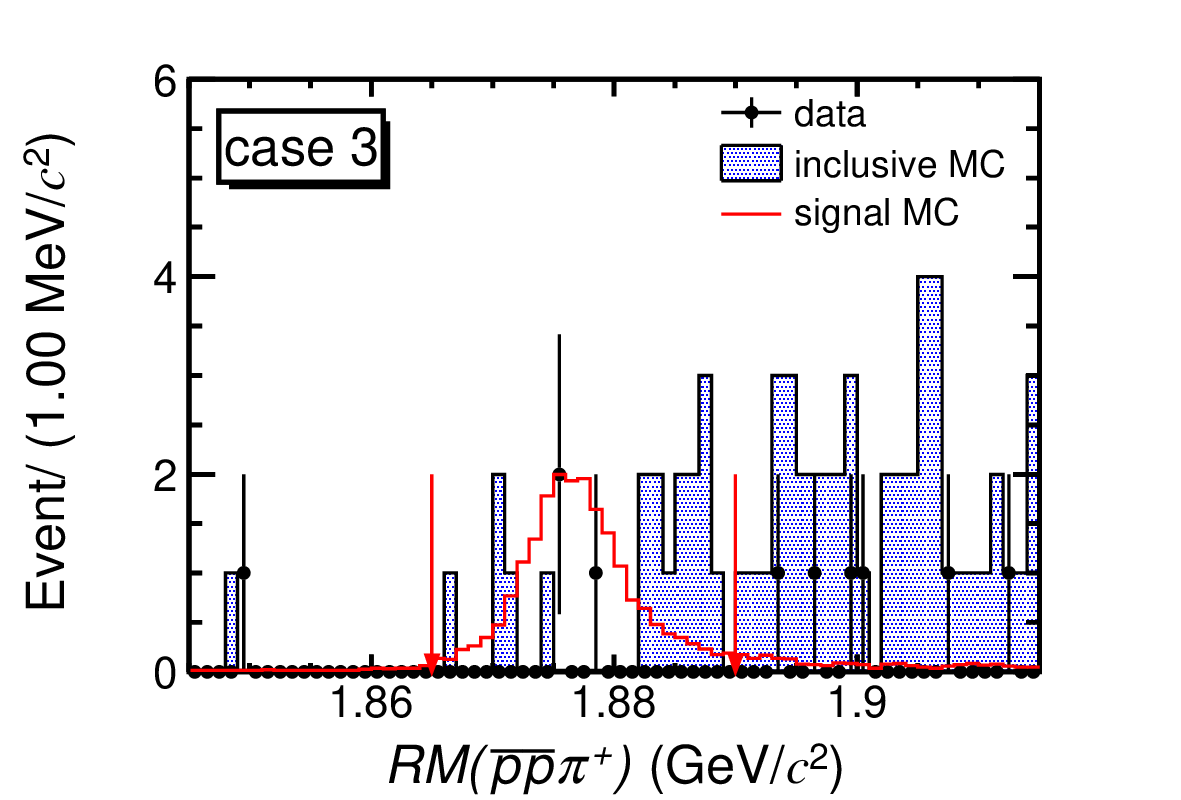}
\includegraphics[angle=0,width=0.43\textwidth]{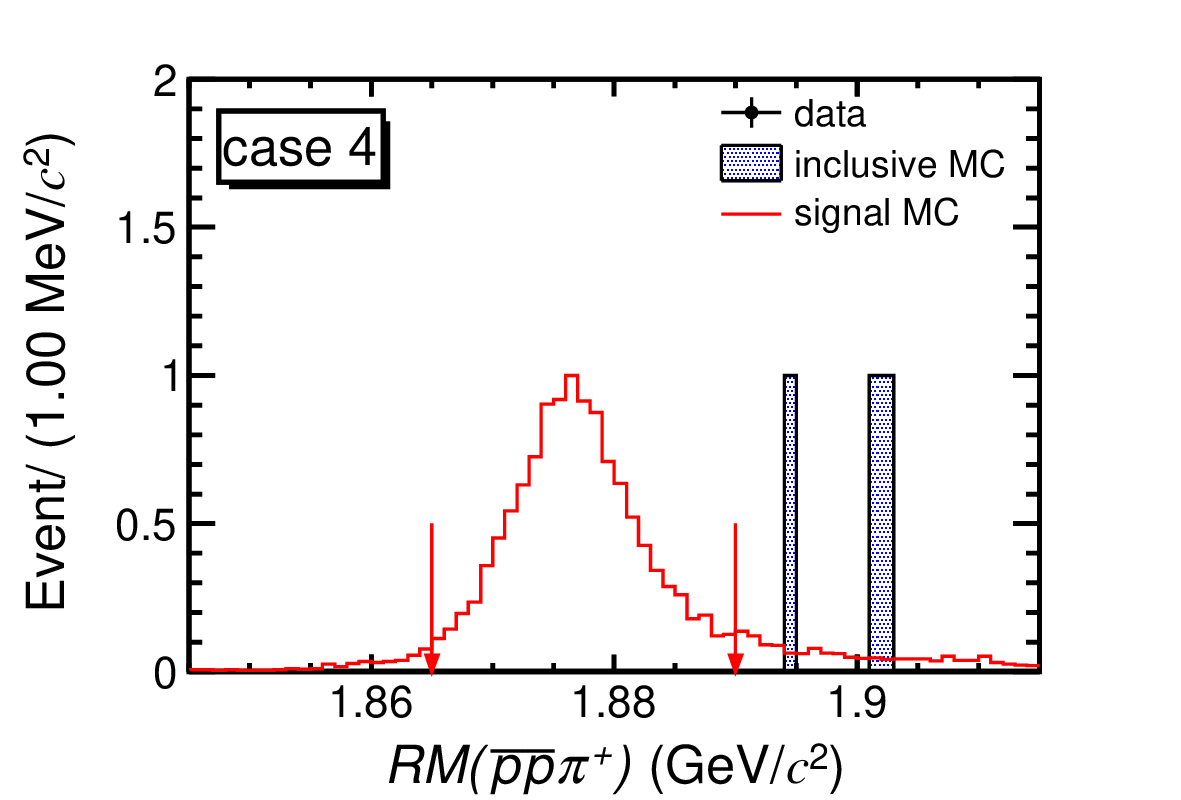}
\caption{Distributions of $RM(pp\pi^-)$ and $RM(\bar{p}\bar{p}\pi^+)$ in data (dots with error bars), inclusive MC (shaded histograms) and the weighted signal MC samples (red histograms) summed over all energy points and after all the selections. The $\bar{d}$ cannot be simulated in the current BESIII software system, so there is no signal MC shape in case 2. The region between the two red arrows is the signal region.}
\label{Fig:reco}
\end{figure*}

To optimize the selection criteria, determine the detection efficiency and estimate the background contributions, Monte Carlo~(MC) simulated samples of events are used. The MC samples are produced with a {\sc geant4}-based~\cite{GEANT4:2002zbu} software package, which includes the geometric description of the BESIII detector and the detector response.
The simulation models the beam energy spread and initial state radiation (ISR) in the $e^+e^-$ annihilations with the generator {\sc kkmc}~\cite{Jadach:2000ir,Jadach:1999vf}. The inclusive MC sample includes the production of open charm processes, the ISR production of vector charmonium-like states, and the continuum processes incorporated in {\sc kkmc}. The known decay modes of charmonium states are modeled with {\sc evtgen}~\cite{Lange:2001uf,Ping:2008zz} using branching fractions taken from the Particle Data Group~\cite{pdg}, and the remaining unknown charmonium decays are modeled with {\sc lundcharm}~\cite{Chen:2000tv,Yang:2014vra}. Final state radiation from charged final state particles is incorporated using the {\sc photos} package~\cite{Richter-Was:1992hxq}.
The signal MC samples of the process $e^+e^-\to pp\pi^-\bar{d}$ together with the charge-conjugate process $e^+e^-\to \bar{p}\bar{p}\pi^+d$ are generated with a phase space model at each $\sqrt{s}$.

\section{data analysis}
To avoid possible bias in the (anti-)deuteron simulation and reconstruction and to improve the selection efficiency, a partial-reconstruction technique is implemented in which only the two protons (antiprotons) and charged $\pi^-$ ($\pi^+$) are reconstructed, and the $\bar{d}$ ($d$) can be missed. Events with three good charged tracks and with a net charge of one and events with four good charged tracks and with zero net charge are selected as candidates. Hereafter, the charge-conjugate mode is always implied, unless explicitly stated otherwise.

For each good charged track, the polar angle $\theta$ is required to be within a range of $|\cos\theta|<0.93$, where $\theta$ is defined with respect to the symmetry axis of the MDC that is taken as the $z$ axis. The distance of closest approach to the interaction point (IP) must be less than 10~cm along the $z$-axis, $|V_{z}|<10$~cm, and less than 1~cm in the transverse plane, $|V_{xy}|<1$~cm.
For particle identification~(PID), the specific energy loss d$E$/d$x$ measured by the MDC and the flight time measured by the TOF are used to form likelihoods $\mathcal{L}(h)~(h=p,K,\pi)$ for each hadron ($h$) hypothesis. Tracks are identified as protons when $\mathcal{L}(p)>0.001$,  $\mathcal{L}(p)>\mathcal{L}(K)$ and $\mathcal{L}(p)>\mathcal{L}(\pi)$, while charged pions are identified when $\mathcal{L}(\pi)>0.001$, $\mathcal{L}(\pi)>\mathcal{L}(p)$ and $\mathcal{L}(\pi)>\mathcal{L}(K)$.
To suppress beam-related background contributions, we require the two protons and the $\pi^-$ to originate from a common vertex and the $\chi^2$ of the vertex fit, $\chi^2_{\rm VF}$, to be less than 70. The vertex position is required to be within a range of $(R_{vx}-0.12$~cm)$^2$+$(R_{vy}+0.14$~cm)$^2\le (0.5$~cm)$^2$ to suppress backgrounds arising from interactions between the beam particles and the beam pipe, where $R_{vx}$ and $R_{vy}$ are the $x$ and $y$ coordinates of the vertex, respectively.

The data samples are divided into four mutually exclusive event classes (`cases') that depend on the reconstruction method. For cases 1 and 2, the signal mode is $e^+e^-\to pp\pi^-\bar{d}$ and the number of reconstructed tracks is 3 or 4, respectively. Similarly, for cases 3 and 4, the signal mode is $e^+e^-\to\bar{p}\bar{p}\pi^+d$ and the number of tracks is 3 or 4, respectively. Different selection criteria are applied to events in the four different classes.

For cases 1 and 3, the deuteron track is not reconstructed, either due to the detector coverage or due to the large ionization of the low momentum track. Based on the simulated MC samples, the transverse momentum recoiling against the $pp\pi^-$ system, $p_{t}$, is required to be less than 0.35~GeV/$c$ when $|\cos\theta_{r}|\le 0.93$, where $\theta_{r}$ is the opening angle between the recoiling system and the $z$-axis.

For cases 2 and 4, the deuteron track is reconstructed, and more information can be used to suppress the background. The mass squared of the deuteron candidate calculated using the TOF information, $m^2_d$, is required to be within a range of $3.2<m^2_d<4.1$~(GeV/$c^2$)$^2$. Here, $m^2_d = p^2 \cdot (\beta^{-2}-1)$, and $\beta = L_{\rm path}/(c \cdot T_{\rm tof})$, where $p$ is the momentum of the deuteron track, $L_{\rm path}$ is the length of the MDC track extrapolated to the
TOF inner radius, $T_{\rm tof}$ is the time of flight corresponding to $L_{\rm path}$ and $c$ is the speed of light. The opening angle $\theta'$ between the candidate deuteron track and the recoil direction of the $pp\pi^-$ system, is required to satisfy $\cos\theta' > 0.9$ to suppress background. Studies based on the inclusive MC samples show that the main surviving background is $e^+e^-\to pp\pi^-\bar{p}\bar{n}+c.c.$ To remove this background process, the number of $\bar{p}$ is required to be zero for case 2 and the number of $p$ is required to be zero for case 4.

\section{background analysis and signal extraction}
\begin{figure}[htbp]
\centering
\subfigure[]{\includegraphics[angle=0,width=0.43\textwidth]{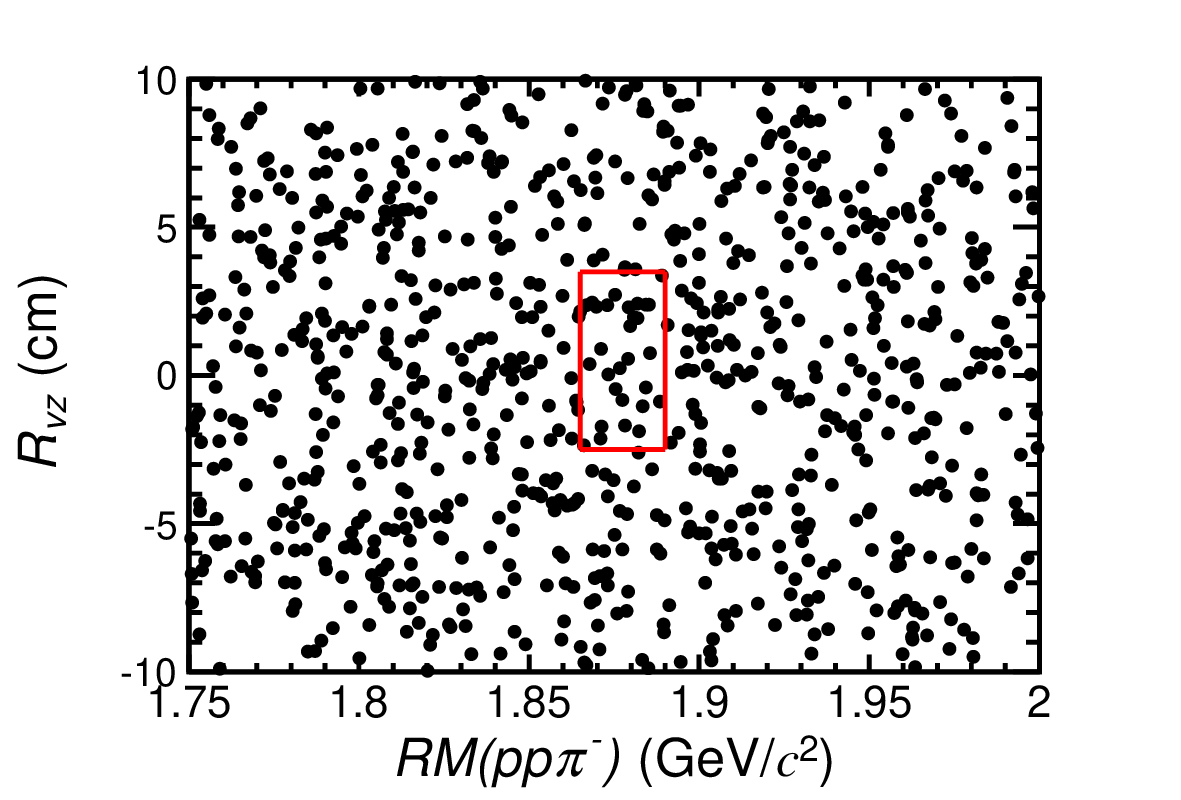}}
\subfigure[]{\includegraphics[angle=0,width=0.43\textwidth]{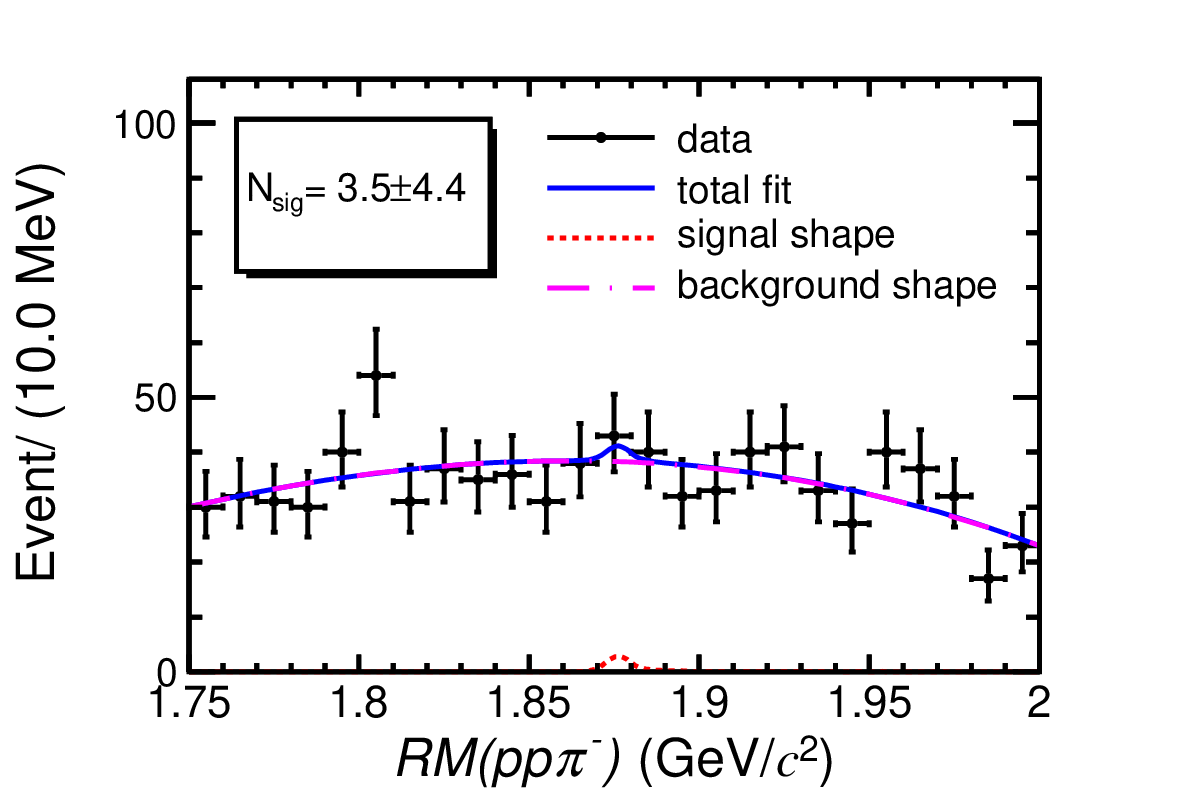}}
\subfigure[]{\includegraphics[angle=0,width=0.43\textwidth]{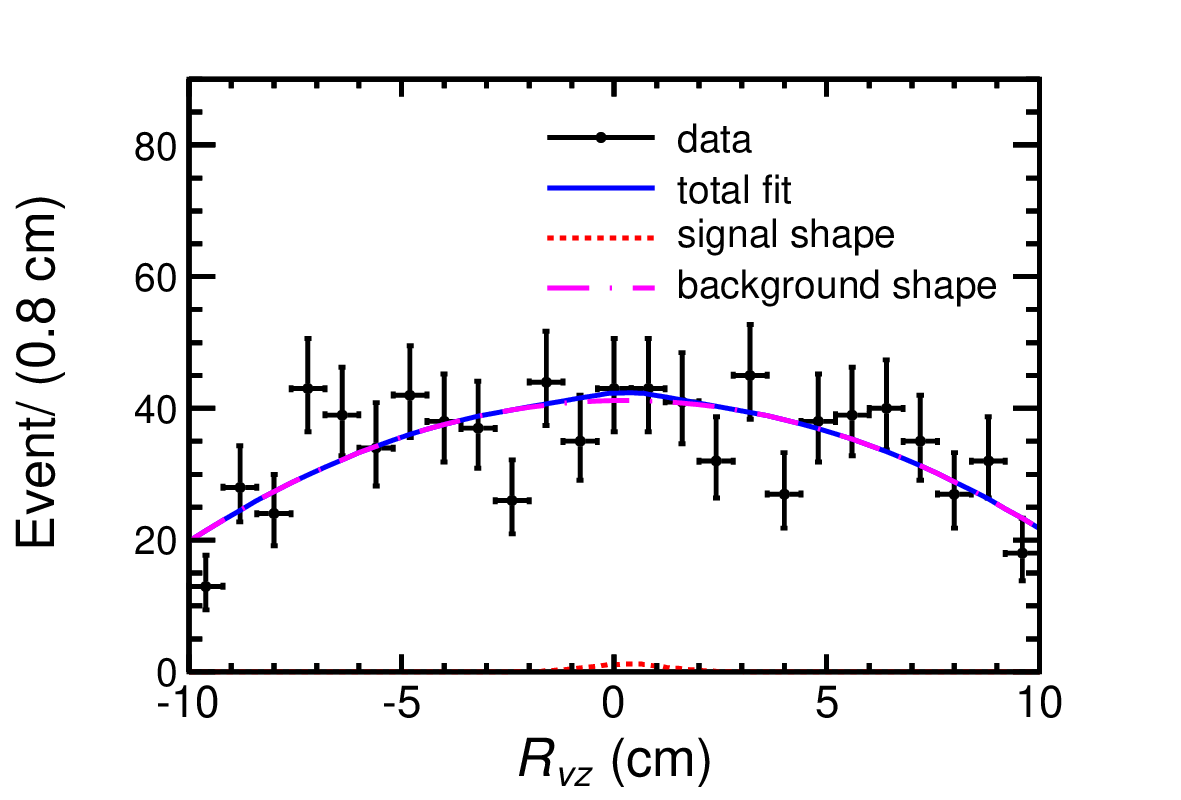}}
\caption{Two-dimensional distribution for $R_{vz}$ versus $RM(pp\pi^-)$ (a) and the projections of the two-dimensional unbinned fit on $RM(pp\pi)$ (b) and $R_{vz}$ (c) in case~1 at $\sqrt{s}=4235.7$ MeV. Here, the red solid box is the signal region and $N_{\rm sig}$ is the signal yield from the fit result.}
\label{Fig:Rvz}
\end{figure}

After imposing all the requirements mentioned above, the distribution of the recoil mass of the $pp\pi^-$ ($RM(pp\pi^-)$) system in data, inclusive MC, and the weighted signal MC samples summed over all energy points is shown in Fig.~\ref{Fig:reco}. For the inclusive MC samples, only a few background events survive, and the remaining events do not form a peak in the $RM(pp\pi^-)$ distribution. Only a few events are present in the data samples, except for case 1. The two-dimensional distribution for $R_{vz}$ and $RM(pp\pi^-)$ in case 1 is shown in Fig.~\ref{Fig:Rvz} at $\sqrt{s}=4235.7$~MeV as an example, where $R_{vz}$ is the $z$ coordinate of the vertex. The flat $R_{vz}$ distribution in data indicates that the main backgrounds in case 1 are not from $\EE$ annihilation, but from the interaction between beam particles and detector material.

\begin{figure}[htbp]
\centering
\subfigure[]{\includegraphics[angle=0,width=0.43\textwidth]{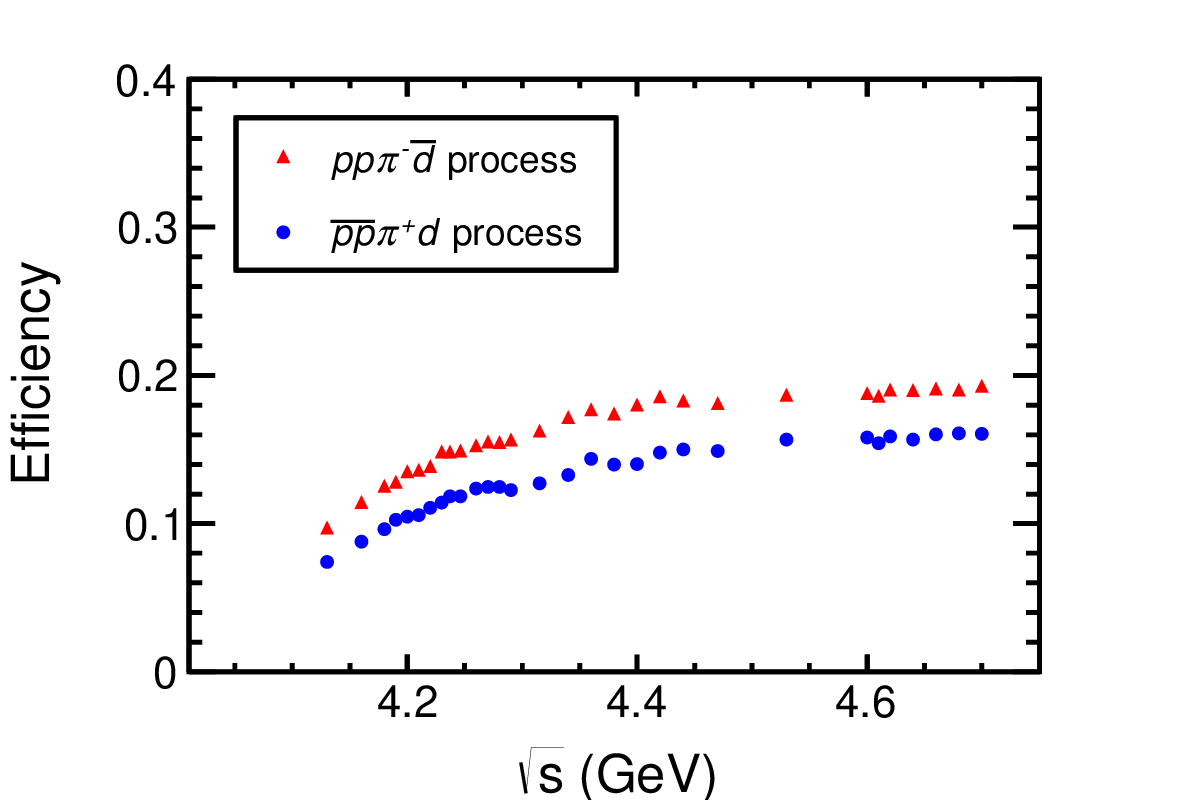}}
\subfigure[]{\includegraphics[angle=0,width=0.43\textwidth]{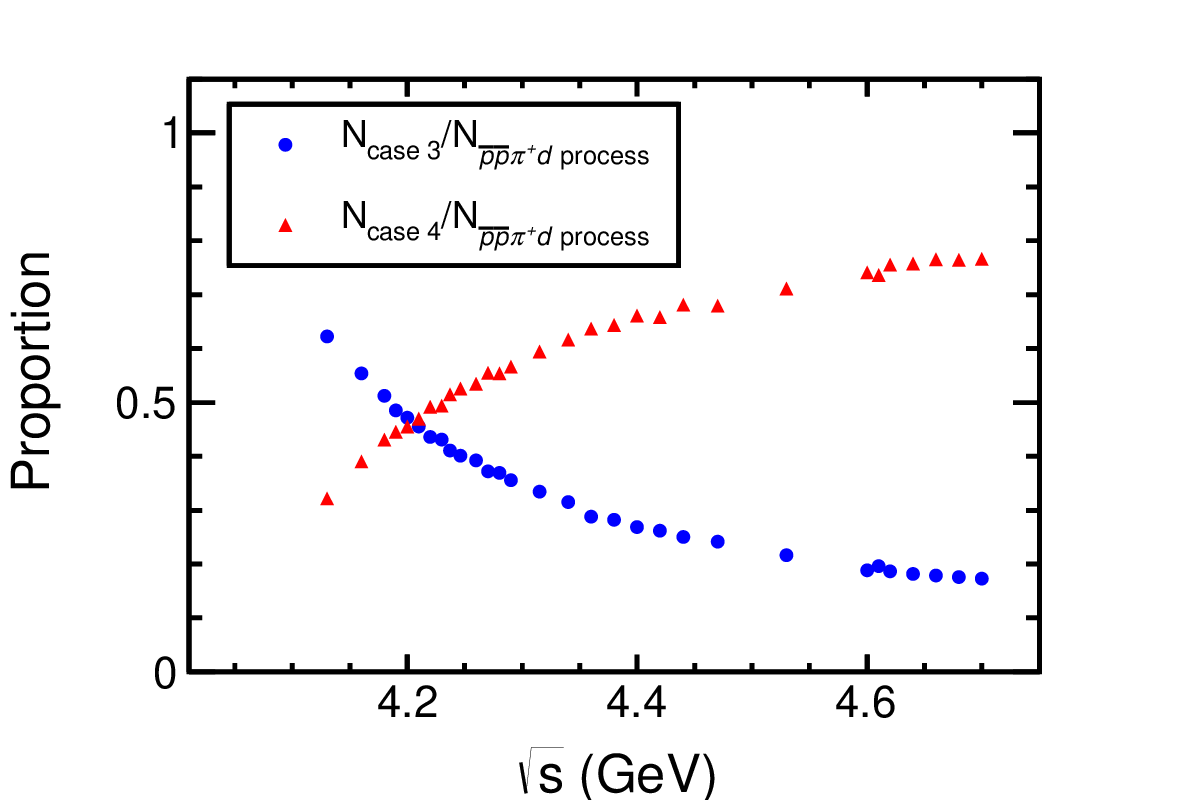}}
\caption{Efficiencies for the $pp\pi^-\bar{d}$ (red triangles) and $\bar{p}\bar{p}\pi^+d$ (blue dots) processes after applying part of the selection criteria as described in Section~\ref{sec:efficiency} (a), and the proportions of case 3 (blue dots) and 4 (red triangles) events in the $\bar{p}\bar{p}\pi^+d$ process (b).}
\label{Fig:eff_pm}
\end{figure}

Different methods are used to determine the signal yields in the different cases. For case 1, a two-dimensional unbinned maximum likelihood fit to the $RM(pp\pi^-)$ versus $R_{vz}$ distribution is performed to determine the signal yields. Here, the signal shape is taken from the signal MC samples, and a first or second order polynomial function is used to describe the background shape. Figure~\ref{Fig:Rvz} shows the fit result projected on $RM(pp\pi^-)$ and $R_{vz}$ at $\sqrt{s}=4235.7$~MeV. For the other cases, a ``counting" method is used to determine the signal yields. The signal events are selected with both $RM(pp\pi^-)$ and $R_{vz}$ within a window of three standard deviations ($\pm3\sigma$) around the mean values, \emph{i.e.}, $1.863<RM(pp\pi^-)<1.892$~GeV/$c^2$ and $-2.5<R_{vz}<3.5$~cm, referred to as the signal region. Here the mean values and standard deviations of the $RM(pp\pi^-)$ and $R_{vz}$ distributions are determined from the signal MC samples. The statistical uncertainty of the number of signal events is estimated with the {\sc trolke}~\cite{Rolke:2004mj,Lundberg:2009iu} package in the {\sc root}~\cite{BRUN199781} framework.

No significant signal is observed from the fit result for case 1, and almost no events survive in the signal region for cases 2, 3 and 4. The signal yields for the different cases are summarized in Table~\ref{Tab:Data_sample}.

\section{Detection efficiency }
\label{sec:efficiency}
The signal MC samples are used to determine the detection efficiency. In the {\sc geant4} software used by the BESIII experiment, the $\bar{d}$ is not simulated. Thus, for cases 1 and 2, which have a $\bar{d}$ in the final state, the detection efficiencies for some of the selection criteria are estimated with control samples selected from data. For cases 3 and 4, the processes can be simulated well and the selection efficiency can be obtained from the signal MC samples directly.

In order to estimate the selection efficiencies for cases 1 and 2, the part of the selection criteria consisting of the requirements on the vertex fit, the vertex position region, and the signal region of $RM(pp\pi^-)$ and $R_{vz}$ are used to select $\bar{p}\bar{p}\pi^+d$ and $pp\pi^-\bar{d}$ processes from the signal MC samples. For each process, the number of tracks can be 3 or 4. Studies based on the high-purity control sample of $e^+e^-\to p\bar{p}\pi^+\pi^-$ show that the proportions of 3 or 4-track events in $\pi^+\pi^-p$ tagged and $\pi^+\pi^-\bar{p}$ tagged samples are very close. So we assume that the proportion of 3 or 4-track events in the $\bar{p}\bar{p}\pi^+d$ process is also close to that in the $pp\pi^-\bar{d}$ process in this analysis. The difference in the proportions between $\pi^+\pi^-p$ tagged and $\pi^+\pi^-\bar{p}$ tagged samples when the number of tracks equals to 3 or 4 in the control sample is taken as the systematic uncertainty, which will be discussed in detail in the next section.
Figure~\ref{Fig:eff_pm}(a) shows the efficiencies of the $pp\pi^-\bar{d}$ and $\bar{p}\bar{p}\pi^+d$ processes after the above selection and Fig.~\ref{Fig:eff_pm}(b) shows the proportions of 3 and 4-track events in the $\bar{p}\bar{p}\pi^+d$ process. The efficiencies of cases 1 and 2 can be calculated with these numbers accordingly.

Table~\ref{Tab:eff} lists the selection efficiencies for the different event classes.  The efficiencies have been corrected for differences between data and MC simulation, as shown by the total correction factors $W$ in this table.  The details on the correction factors are discussed in the next section.

\begin{table*}[htbp]
	\setlength{\abovecaptionskip}{0.cm}
	\setlength{\belowcaptionskip}{2pt}
	\centering
	\caption{The total correction factor $w$ and the corresponding corrected efficiency $\epsilon$.}
	\normalsize  							
	\setlength{\tabcolsep}{12pt}			
	\renewcommand{\arraystretch}{1.2}  		
	\begin{tabular}{ccccccccc}
        \hline
		\multirow{2}{*}{$\sqrt{s}$ (MeV)}   & \multicolumn{2}{c}{case 1} & \multicolumn{2}{c}{case 2} & \multicolumn{2}{c}{case 3} & \multicolumn{2}{c}{case 4} \\
        \cline{2-9}
		& $W$ & $\epsilon$ & $W$ & $\epsilon$ & $W$ & $\epsilon$ & $W$ & $\epsilon$   \\
        \hline
	    4128.5 & 0.925 &  0.056 & 1.036 & 0.032  & 1.023 & 0.047  & 1.034 & 0.025  \\
	    4157.4 & 0.925 &  0.059 & 1.035 & 0.046  & 1.021 & 0.050  & 1.031 & 0.035  \\
	    4178.0 & 0.927 &  0.060 & 1.035 & 0.056  & 1.019 & 0.050  & 1.027 & 0.043  \\
	    4188.8 & 0.927 &  0.058 & 1.035 & 0.059  & 1.018 & 0.051  & 1.027 & 0.047  \\
	    4198.9 & 0.927 &  0.059 & 1.034 & 0.064  & 1.019 & 0.050  & 1.025 & 0.049  \\
	    4209.2 & 0.927 &  0.058 & 1.033 & 0.066  & 1.018 & 0.049  & 1.024 & 0.051  \\
	    4218.7 & 0.928 &  0.056 & 1.034 & 0.071  & 1.018 & 0.049  & 1.023 & 0.056  \\
	    4226.3 & 0.929 &  0.059 & 1.033 & 0.076  & 1.018 & 0.050  & 1.022 & 0.058  \\
	    4235.7 & 0.928 &  0.057 & 1.034 & 0.079  & 1.017 & 0.049  & 1.023 & 0.062  \\
	    4243.8 & 0.928 &  0.056 & 1.034 & 0.081  & 1.017 & 0.048  & 1.022 & 0.064  \\
	    4258.0 & 0.929 &  0.056 & 1.034 & 0.084  & 1.017 & 0.049  & 1.022 & 0.068  \\
	    4266.8 & 0.928 &  0.054 & 1.032 & 0.089  & 1.016 & 0.047  & 1.020 & 0.071  \\
	    4277.7 & 0.928 &  0.053 & 1.033 & 0.088  & 1.016 & 0.047  & 1.021 & 0.070  \\
	    4287.9 & 0.929 &  0.052 & 1.033 & 0.092  & 1.016 & 0.044  & 1.020 & 0.071  \\
	    4312.0 & 0.927 &  0.050 & 1.033 & 0.010  & 1.013 & 0.043  & 1.019 & 0.077  \\
	    4337.4 & 0.930 &  0.050 & 1.033 & 0.110  & 1.015 & 0.042  & 1.017 & 0.083  \\
	    4358.3 & 0.931 &  0.048 & 1.033 & 0.116  & 1.015 & 0.042  & 1.017 & 0.093  \\
	    4377.4 & 0.930 &  0.046 & 1.033 & 0.116  & 1.014 & 0.040  & 1.016 & 0.092  \\
	    4396.4 & 0.930 &  0.045 & 1.032 & 0.123  & 1.013 & 0.038  & 1.015 & 0.094  \\
	    4415.6 & 0.931 &  0.045 & 1.032 & 0.126  & 1.013 & 0.039  & 1.014 & 0.099  \\
	    4436.2 & 0.931 &  0.043 & 1.032 & 0.129  & 1.013 & 0.038  & 1.014 & 0.104  \\
	    4467.1 & 0.932 &  0.041 & 1.032 & 0.127  & 1.013 & 0.036  & 1.013 & 0.103  \\
	    4527.1 & 0.931 &  0.038 & 1.032 & 0.137  & 1.012 & 0.034  & 1.013 & 0.113  \\
	    4599.5 & 0.931 &  0.032 & 1.033 & 0.144  & 1.010 & 0.030  & 1.012 & 0.118  \\
	    4611.9 & 0.931 &  0.034 & 1.033 & 0.142  & 1.010 & 0.030  & 1.012 & 0.115  \\
	    4628.0 & 0.931 &  0.033 & 1.032 & 0.148  & 1.010 & 0.030  & 1.011 & 0.121  \\
	    4640.9 & 0.932 &  0.032 & 1.033 & 0.149  & 1.010 & 0.029  & 1.011 & 0.120  \\
	    4661.2 & 0.931 &  0.032 & 1.032 & 0.151  & 1.009 & 0.029  & 1.011 & 0.124  \\
	    4681.9 & 0.931 &  0.031 & 1.032 & 0.150  & 1.008 & 0.028  & 1.010 & 0.124  \\
	    4698.8 & 0.931 &  0.031 & 1.032 & 0.152  & 1.009 & 0.028  & 1.010 & 0.124  \\
   \hline
	\end{tabular}
	\label{Tab:eff}
\end{table*}

\section{systematic uncertainty}
The systematic uncertainties in the cross section measurement originate from the integrated luminosity, the tracking and PID efficiencies, and the determination of the signal yields, which includes the fit range, the signal and background shape, etc. These systematic uncertainties are listed in Table~\ref{Tab:sys}
and discussed in more detail below.

\begin{table*}[htbp]
\setlength{\abovecaptionskip}{0.cm}
\setlength{\belowcaptionskip}{2pt}
\centering
\caption{Relative systematic uncertainties (in units of \%) in the cross section measurement of $e^+e^-\to pp\pi^-\bar{d}+c.c.$}
\normalsize								
\setlength{\tabcolsep}{12pt}			
\renewcommand{\arraystretch}{1.2}		
	\begin{tabular}{ccccc}
	\hline
		Source                            &    case 1    &  case 2     &  case 3      & case 4         \\
		\hline
		Luminosity                        &     1.0      &   1.0       &     1.0      &   1.0          \\
		Tracking and PID efficiencies       &   0.5-1.4    &   0.7-1.9   &    0.6-1.5   &  0.8-1.9       \\
		Vertex fit                        &     0.6      &     0.6     &     0.6      &    0.6         \\
		$R_{vx}$ and $R_{vy}$             &    0.3       &     0.3     &     0.3      &    0.3         \\
		$p_{t}$ and $\cos\theta_{r}$      &    2.2-3.1   &    -        &   2.2-3.1    &     -          \\
		Open angle $\cos\theta'$          &     -        &     -       &     -        &     -          \\
		$m^2_d$                           &     -        &    0.6-6.0  &     -        &   0.6-6.0      \\
		$p(\bar{p})$ veto                 &     -        &    0.4-2.1  &     -        &   0.4-2.1      \\
		Efficiency estimation             &    3.0       &    1.0      &     -        &     -          \\
		Signal yield extraction              &   17.7       &    11.8     &    11.8      &     11.8       \\
		Sum                               &  18.1-18.3   &   12.0-13.4 &   12.1-12.3  &   11.9-13.4    \\
	 \hline
	 \end{tabular}
	 \label{Tab:sys}
\end{table*}

The integrated luminosity is measured using Bhabha scattering events with an uncertainty of 1.0\%~\cite{BESIII:2015qfd,BESIII:2022ulv,BESIII:2022dxl}.

The tracking and PID efficiencies are studied with a high-purity control sample of $J/\psi\to p\bar{p}\pi^+\pi^-$ events, and the polar angle and transverse momentum ($p_{t}$) dependent efficiencies are measured~\cite{BESIII:2020lkm}. The efficiency of MC events is corrected by the two-dimensional efficiency scale factors, and the uncertainty is estimated by varying the efficiency scale factors by one standard deviation for each $p_{t}$ versus $\cos\theta$ bin. The differences between the new efficiencies and the nominal ones are taken as the systematic uncertainties.

In this analysis, the selection efficiency is corrected according to the measurements with control samples selected from data directly. The efficiency correction factor $w^{\rm s}$ is defined as
\begin{equation}
   w^{\rm s}=\epsilon_{\rm data}^{\rm s}/\epsilon_{{\rm MC}}^{\rm s},
\end{equation}
where the subscripts ``MC" and ``data" represent MC simulation and data samples, respectively, the superscript ``s" represents the selection criterion ``s" and $\epsilon^{\rm s}$ is the efficiency for the selection ``s" which is calculated as
\begin{equation}
        \epsilon^{\rm s}=\frac {n_{\rm sig}^{\rm s}} {n_{\rm sig}^{\rm s}+n_{\rm bkg}^{\rm s}},
\end{equation}
for both data and MC samples. Here $n_{\rm sig}^{\rm s}$ is the number of events in the signal region with selection criterion ``s" and $n_{\rm bkg}^{\rm s}$ is the number of events out of the region of selection criterion ``s". The uncertainty of $\epsilon^{\rm s}$ and the uncertainty of $w^{\rm s}$ is calculated as
\begin{equation}
        \sigma_{\epsilon^{\rm s}}=\frac {n_{\rm bkg}^{\rm s}} {n_{\rm sig}^{\rm s}} \cdot \frac {\sqrt{ {(\frac {\sigma_{n_{\rm sig}^{\rm s}}} {n_{\rm bkg}^{\rm s}})}^2+ {{(\frac {\sigma_{n_{\rm bkg}^{\rm s}}} {n_{\rm bkg}^{\rm s}})}^2 }}} {{(1+\frac {n_{\rm bkg}^{\rm s}} {n_{\rm sig}^{\rm s}})}^2},
\end{equation}
\begin{equation}
    \sigma_{w^{\rm s}}=\frac {\epsilon_{\rm data}^{\rm s}}{\epsilon_{\rm MC}^{\rm s}}
    \cdot \sqrt{ {(\frac {\sigma_{\epsilon_{\rm data}^{\rm s}}}{\epsilon_{\rm data}^{\rm s}})}^2+ {(\frac {\sigma_{\epsilon_{\rm MC}^{\rm s}}}{\epsilon_{\rm MC}^{\rm s}})}^2},
\end{equation}
where $\sigma_{n_{\rm sig}^{\rm s}}$ and $\sigma_{n_{\rm bkg}^{\rm s}}$ are the uncertainties of $n_{\rm sig}^{\rm s}$ and $n_{\rm bkg}^{\rm s}$. For $w^{\rm s}=(1+\Delta w^{\rm s})\pm\sigma_{w^{\rm s}}$, if $|\frac {\Delta w^{\rm s}} {\sigma_{w^{\rm s}}}|<1.0$, the MC simulation is consistent with the data, no correction will be applied, and $|\Delta w^{\rm s}|+\sigma_{w^{\rm s}}$ is taken as the systematic uncertainty. On the other hand, if $|\frac {\Delta w^{\rm s}} {\sigma_{w^{\rm s}}}|>1.0$, the MC efficiency will be corrected to data, and the new efficiency after correction is $\epsilon=\epsilon_{\rm MC}\cdot w^{\rm s}$, and $\sigma_{w^{\rm s}}$ is taken as the systematic uncertainty. The total correction factor $W$ is defined as $\prod\limits_{\rm s}\limits^{}{w^{\rm s}}$, which is shown in Table~\ref{Tab:eff}.

The uncertainties of the vertex fit and vertex position are studied using a control sample of $e^+e^-\to\pi^+\pi^-J/\psi$, $J/\psi\to e^+e^-/\mu^+\mu^-$ events. The efficiency ratio between data and MC simulation with the vertex fit requirement $\chi_{\rm VF}^2<70$ is $w^{\rm VF}=1.002\pm 0.004$, and we take $\sigma^{\rm VF}=0.006$ as the systematic uncertainty and set $w^{\rm VF}=1.0$. The efficiency ratio between data and MC simulation with the vertex position requirement $(R_{vx}-0.12$~cm)$^2$+$(R_{vy}+0.14$~cm)$^2\le (0.5$~cm)$^2$ is $w^{\rm VP}=1.001\pm 0.002$, and we take $\sigma^{\rm VP}=0.003$ as the systematic uncertainty and set $w^{\rm VP}=1.0$.

The systematic uncertainty due to the $p_{t}$ and $\cos\theta_{r}$ requirements is estimated by loosening or tightening the nominal requirements. The transverse momentum of the track recoiling against the $pp\pi^-$ system, $p_{t}$, is required to be less than 0.33 or 0.37~GeV/$c$ when $|\cos\theta_{r}|\le 0.93$. The largest changes of the efficiency compared to the nominal requirement range from 2.2\% to 3.1\% and are taken as the corresponding uncertainties.

Because we do not have a pure deuteron sample, due to the different time resolution of the TOF between data and MC simulation, the uncertainty from the $m^2_d$ requirement is studied with the control sample $e^+e^-\to p\bar{p}\pi^+\pi^-$. The $m^2$ requirement for the recoil mass of $\pi^+\pi^-\bar{p}$ is applied at each energy point in the control sample. Different $m^2$ ranges are chosen at each energy point to ensure that the efficiency in the control sample is the same as that in this analysis. The efficiency difference between data and MC simulation, which ranges from 0.6\% to 6.0\%, is taken as the systematic uncertainty.

The systematic uncertainty of the $p(\bar{p})$ veto in the $\bar{p}\bar{p}\pi^+d(pp\pi^-\bar{d})$ process is due to the difference in proportion of deuterons misidentified as protons between data and signal MC samples. The proportion of deuterons misidentified as protons in the simulation can be calculated from the signal MC sample, which is taken as the systematic uncertainty directly since the difference between data and MC sample is small. The corresponding uncertainty ranges from 0.4\% to 2.1\%.

The efficiency of the $\cos\theta'$ selection is very high due to the loose requirement, and its systematic uncertainty is negligible.

Since the $\bar{d}$ cannot be simulated, the $\bar{p}\bar{p}\pi^+d$ process is used to estimate the efficiency in the $pp\pi^-\bar{d}$ process, which is described in detail in the previous section. To estimate the systematic uncertainty, we take $e^+e^-\to p\bar{p}\pi^+\pi^-$ as the control sample. The proportions of 3-track or 4-track events in the $\pi^+\pi^-\bar{p}$ tagged and $\pi^+\pi^-p$ tagged samples are obtained from the control sample and the difference in proportions between $\pi^+\pi^-p$ tagged and $\pi^+\pi^-\bar{p}$ tagged samples when the number of tracks equals 3 or 4 is taken as the systematic uncertainty. Based on studies of the control sample, the correction factor and the systematic uncertainty are calculated, where the correction factors are 0.93 and 1.03 for cases 1 and 2, respectively, and the systematic uncertainties are 3\% and 1\% for cases 1 and 2, respectively.

For the uncertainty of the signal yield extraction, alternative methods are used to estimate the signal yield. For case 1, the uncertainty of the signal yield extraction is estimated by varying the fit range, and changing the signal and background shape. We vary the upper and lower bounds of $RM(pp\pi)$ and $R_{vz}$ by $\pm 5$~MeV/$c^2$ and $\pm 1$~cm, respectively, use a Gaussian function to describe the signal shape, and use a third-order polynomial function to describe the background shape. The largest difference of the cross section compared with the nominal one is taken as the systematic uncertainty, which is 17.7\%. For cases 2, 3 and 4, the control samples, $e^+e^-\to p\bar{p}\pi^+\pi^-$ and $J/\psi\to p\bar{p}\pi^+\pi^-$, are used to estimate the systematic uncertainties of the $R_{vz}$ and $RM(pp\pi^-)$ window selection, respectively. The largest difference in efficiencies between data and simulated MC samples is taken as the systematic uncertainty, which is 11.8\%.

The total systematic uncertainties for the cross section measurement in different cases are summarized in Table~\ref{Tab:sys}. Totals are obtained by summing the individual values in quadrature under the assumption that all the sources are independent.

\section{Upper Limit of the Cross Section}
The upper limit of the cross section at the $90\%$ confidence level (C.L.), taking into account the systematic uncertainty, is calculated using a Bayesian method. For case 1, a scan of the likelihood distribution ($\mathcal{L}$) as a function of the signal yield ($N_{\rm sig}$) is obtained from fits to $RM(pp\pi^-)$ versus $R_{vz}$ with fixed values for the signal yield. To take the systematic uncertainty into consideration, the likelihood distribution is convolved with a Gaussian function with mean value zero and standard deviation $N_{\rm sig}\cdot\sigma_{\rm sys}$. The upper limit on the signal yield ($N_{\rm sig}^{\rm ul}$) at the 90\% C.L. is determined as
 $ {\int_{0}^{N_{\rm sig}^{\rm ul}}\mathcal{L}\,dN_{\rm sig}}
  /{\int_{0}^{\infty}\mathcal{L}\,dN_{\rm sig}}$.
For cases 2, 3, and 4, a likelihood function is constructed to calculate the signal yield at the $90\%$ C.L.\ assuming that the numbers of signal ($N_{s}$) and background ($N_{b}$) events obey a Poisson distribution, and the efficiency ($\epsilon$) obeys a Gaussian distribution
 \begin{equation}
    \begin{aligned}
        &P(N_{s})=Pois(N_{s};\epsilon\mu),\\
        &P(N_{b})=Pois(N_{b};\tau b),\\
        &P(\epsilon)=G(\epsilon;z,\sigma_{\epsilon}),
    \end{aligned}
 \end{equation}
where $\mu$ is the signal rate in the signal region, $b$ is the number of background events in the signal region, $z$ is the efficiency obtained from MC simulation, $\sigma_{\epsilon}$ is the absolute systematic uncertainty, $\sigma_{\epsilon}= \epsilon\cdot \sigma_{\rm sys}$, where $\sigma_{\rm sys}$ is the relative total systematic uncertainty, and $\tau$ is the ratio of the sizes of the sideband regions and the signal regions.

\begin{figure}[htbp]
\centering
\includegraphics[angle=0,width=0.43\textwidth]{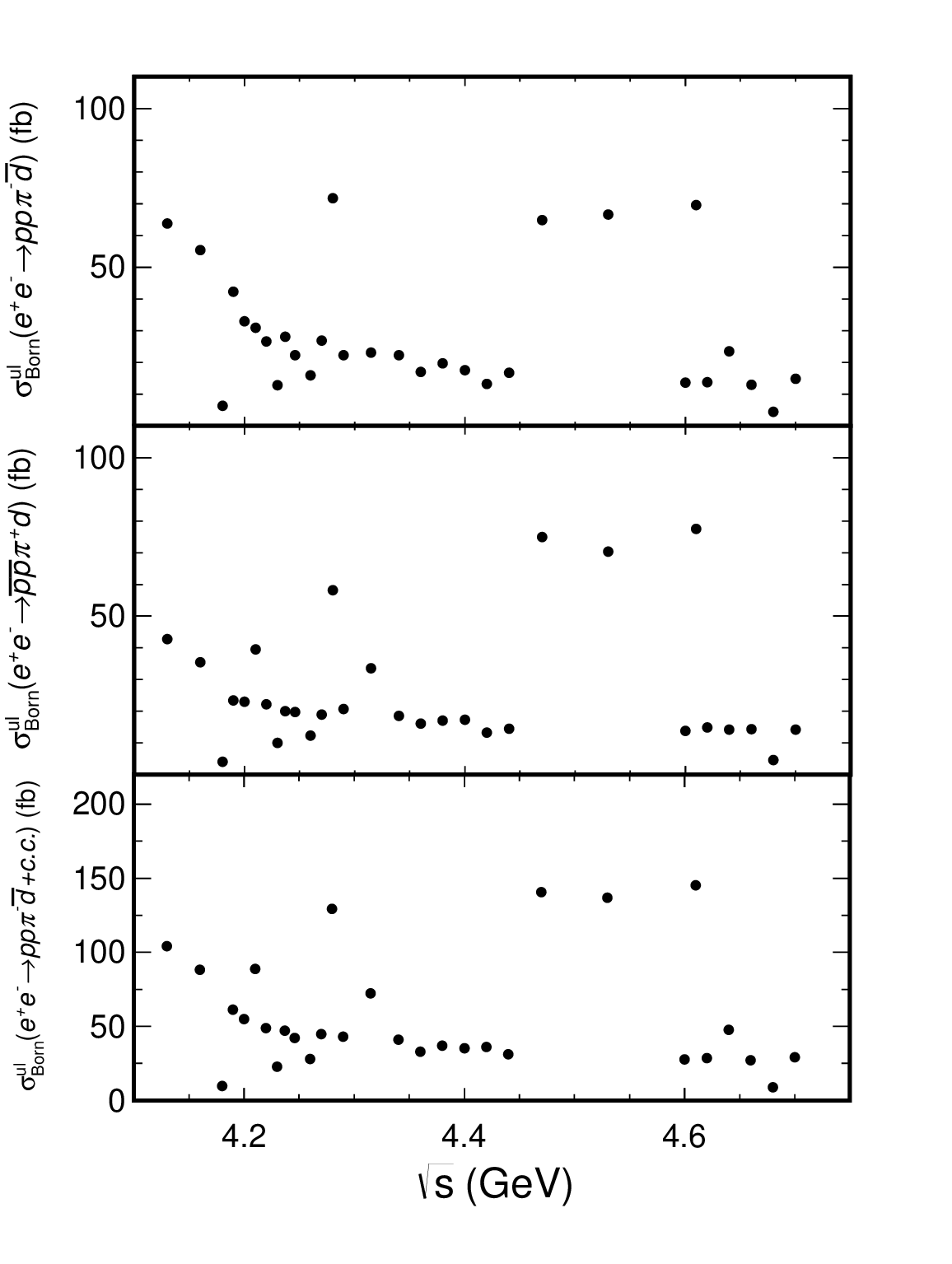}
\caption{Upper limits of the Born cross sections at the $90\%$ C.L.}
\label{Fig:born_cs}
\end{figure}

Since there are very few events in the sideband region, we fix $b$ to 0 in the likelihood function to get a more conservative upper limit. The likelihood function is defined as
 \begin{equation}
        L(\mu,\epsilon\ |\ N_{s},z)=\frac{(\epsilon\mu)^{N_s}}{N_s!}e^{-\epsilon\mu}
        \times\frac{1}{\sqrt{2\pi}\sigma_{\epsilon}}
        e^{-\frac{{(\epsilon-z)}^2}{2\sigma_{\epsilon}^2}}.
        \label{equ:upper}
 \end{equation}
 The signal yield at the $90\%$ C.L.\ is determined using Eq.~(\ref{equ:upper}). Since the data samples for the four cases are completely independent, the likelihood functions for the different cases can be multiplied to get a combined likelihood function. The signal yields at the $90\%$ C.L.\ for the processes of $e^+e^-\to pp\pi^-\bar{d}$, $e^+e^-\to \bar{p}\bar{p}\pi^+d$, and $e^+e^-\to pp\pi^-\bar{d}+c.c.$, which with the combination of cases 1 and 2, cases 3 and 4, and all four cases, are calculated. The corresponding upper limit of the Born cross section at the $90\%$ C.L. is calculated as
 \begin{equation}
       \sigma^{\rm ul}=
       \frac{N^{\rm ul}_{\rm sig}}{\mathcal{L}\cdot\epsilon\cdot(1+\delta)\cdot
       \frac{1}{|1-\Pi|^2}},
 \end{equation}
where $N^{\rm ul}_{\rm sig}$ is the upper limit of the signal yield at the $90\%$ C.L., $\mathcal{L}$ is the integrated luminosity of the data set, $\epsilon$ is the detection efficiency, $(1+\delta)$ and $\frac{1}{|1-\Pi|^2}$ are the ISR and vacuum polarization~\cite{vacuum} correction factors, respectively. We use a flat dressed cross section line shape as the input to generate the signal MC sample and obtain the corresponding ISR correction factor $(1+\delta)$, since no significant signal is seen in data. The numbers related to the Born cross section measurement are summarized in Table~\ref{Tab:detail}.  Figure~\ref{Fig:born_cs} shows the upper limits of the Born cross sections at the 90\% C.L. for the processes of $e^+e^-\to pp\pi^-\bar{d}$, $e^+e^-\to \bar{p}\bar{p}\pi^+d$, and $e^+e^-\to pp\pi^-\bar{d}+c.c.$

\section{summary}
We search for the process $e^+e^-\to pp\pi^-\bar{d}+c.c.$ with the BESIII data at $\sqrt{s}$ from 4.13 to 4.70~GeV. No significant signal is observed. The upper limits at the $90\%$ C.L on the Born cross sections of $e^+e^-\to pp\pi^-\bar{d}$, $\bar{p}\bar{p}\pi^+d$ and $pp\pi^-\bar{d}+c.c.$ are determined to be from 4.3 to 72~fb, 4.1 to 76~fb, and 9.0 to 145~fb, respectively. The current BESIII sensitivity is in the same order of magnitude as the cross section of inclusive $\bar{d}$ production $\sigma(e^+e^-\to\bar{d}X)=(9.63\pm 0.41(\rm stat)^{+1.17}_{-1.01}(\rm syst))$~fb at $\sqrt{s}\approx 10.58$~GeV from the BaBar experiment~\cite{BaBar:2014ssg}.

In the BESIII experiment, the low backgrounds guarantee a high sensitivity measurement of the cross section in the process $e^+e^-\to\bar{p}\bar{p}\pi^+d$ with the number of tracks equal to 3 or 4 and $e^+e^-\to pp\pi^-\bar{d}$ with the number of tracks equal to 3, and an improved sensitivity may be obtained if the beam backgrounds can be further suppressed in the process $e^+e^-\to pp\pi^-\bar{d}$ with the number of tracks equal to 4. BEPCII is upgrading the luminosity performance and increasing the maximum $\sqrt{s}$~\cite{BESIII:2020nme}, which will enable a further search for the (anti)deuteron, the $d^*(2380)$, and other possible states with six quarks at the BESIII experiment.

\section{ACKNOWLEDGEMENT}
The BESIII collaboration thanks the staff of BEPCII and the IHEP computing center for their strong support. This work is supported in part by National Key R\&D Program of China under Contracts Nos. 2020YFA0406300, 2020YFA0406400; National Natural Science Foundation of China (NSFC) under Contracts Nos. 11635010, 11735014, 11835012, 11935015, 11935016, 11935018, 11961141012, 12022510, 12025502, 12035009, 12035013, 12061131003, 12192260, 12192261, 12192262, 12192263, 12192264, 12192265; the Chinese Academy of Sciences (CAS) Large-Scale Scientific Facility Program; the CAS Center for Excellence in Particle Physics (CCEPP); Joint Large-Scale Scientific Facility Funds of the NSFC and CAS under Contract No. U1832207; CAS Key Research Program of Frontier Sciences under Contracts Nos. QYZDJ-SSW-SLH003, QYZDJ-SSW-SLH040; 100 Talents Program of CAS; The Institute of Nuclear and Particle Physics (INPAC) and Shanghai Key Laboratory for Particle Physics and Cosmology; ERC under Contract No. 758462; European Union's Horizon 2020 research and innovation programme under Marie Sklodowska-Curie grant agreement under Contract No. 894790; German Research Foundation DFG under Contracts Nos. 443159800, 455635585, Collaborative Research Center CRC 1044, FOR5327, GRK 2149; Istituto Nazionale di Fisica Nucleare, Italy; Ministry of Development of Turkey under Contract No. DPT2006K-120470; National Research Foundation of Korea under Contract No. NRF-2022R1A2C1092335; National Science and Technology fund; National Science Research and Innovation Fund (NSRF) via the Program Management Unit for Human Resources \& Institutional Development, Research and Innovation under Contract No. B16F640076; Polish National Science Centre under Contract No. 2019/35/O/ST2/02907; Suranaree University of Technology (SUT), Thailand Science Research and Innovation (TSRI), and National Science Research and Innovation Fund (NSRF) under Contract No. 160355; The Royal Society, UK under Contract No. DH160214; The Swedish Research Council; U. S. Department of Energy under Contract No. DE-FG02-05ER41374.

\begin{table*}[h!]
	\setlength{\abovecaptionskip}{0.cm}
	\setlength{\belowcaptionskip}{2pt}
	\centering
	\caption{The upper limit of the Born cross section at the 90\% C.L. and the numbers used in the measurement.}
	\normalsize  							
	\setlength{\tabcolsep}{12pt}			
	\renewcommand{\arraystretch}{1.2}  		
	\begin{tabular}{cccccc}
        \hline
        $\sqrt{s}$ (MeV) & $1+\delta$   & $\frac{1}{|1-\Pi|^2}$  & $\sigma^{\rm ul}(pp\pi^-\bar{d})$ (fb)		   &$\sigma^{\rm ul}( \bar{p}\bar{p}\pi^+d)$ (fb) &  $\sigma^{\rm ul}(pp\pi^-\bar{d}+c.c.)$ (fb)  \\
        \hline
	    4128.5  &    0.886   &    1.052 &     64    &    43    &   104   \\
	    4157.4  &    0.885   &    1.053 &     55    &    35    &   88   \\
	    4178.0  &    0.897   &    1.054 &    6.3    &   4.1    &  9.9   \\
	    4188.8  &    0.898   &    1.056 &     42    &    23    &   61   \\
	    4198.9  &    0.900   &    1.056 &     33    &    23    &   55   \\
	    4209.2  &    0.903   &    1.057 &     31    &    40    &   89   \\
	    4218.7  &    0.904   &    1.056 &     27    &    22    &   49   \\
	    4226.3  &    0.905   &    1.056 &     13    &    10    &   23   \\
	    4235.7  &    0.908   &    1.056 &     28    &    20    &   47   \\
	    4243.8  &    0.907   &    1.056 &     22    &    20    &   42   \\
	    4258.0  &    0.909   &    1.054 &     16    &    12    &   28   \\
	    4266.8  &    0.912   &    1.053 &     27    &    19    &   45   \\
	    4277.7  &    0.912   &    1.053 &     72    &    58    &   129   \\
	    4287.9  &    0.915   &    1.053 &     22    &    21    &   43   \\
	    4312.0  &    0.917   &    1.052 &     23    &    34    &   72   \\
	    4337.4  &    0.921   &    1.051 &     22    &    19    &   41   \\
	    4358.3  &    0.925   &    1.051 &     17    &    16    &   33   \\
	    4377.4  &    0.925   &    1.051 &     20    &    17    &   37   \\
	    4396.4  &    0.927   &    1.051 &     18    &    17    &   35   \\
	    4415.6  &    0.928   &    1.052 &     13    &    13    &   36   \\
	    4436.2  &    0.932   &    1.054 &     17    &    14    &   31   \\
	    4467.1  &    0.934   &    1.055 &     65    &    75    &   141   \\
	    4527.1  &    0.939   &    1.054 &     67    &    70    &   137   \\
	    4599.5  &    0.896   &    1.055 &     14    &    14    &   28   \\
	    4611.9  &    0.943   &    1.054 &     69    &    76    &   145   \\
	    4628.0  &    0.945   &    1.054 &     13    &    15    &   29   \\
	    4640.9  &    0.946   &    1.054 &     23    &    14    &   48   \\
	    4661.2  &    0.947   &    1.054 &     13    &    14    &   27   \\
	    4681.9  &    0.948   &    1.054 &    4.3    &   4.5    &  9.0   \\
	    4698.8  &    0.949   &    1.054 &     14    &    14    &   29   \\
   \hline
	\end{tabular}
	\label{Tab:detail}
\end{table*}

\clearpage
\nocite{*}
\bibliographystyle{apsrev4-2}
\bibliography{refer}

\begin{thebibliography}{57}%
\makeatletter
\providecommand \@ifxundefined [1]{%
 \@ifx{#1\undefined}
}%
\providecommand \@ifnum [1]{%
 \ifnum #1\expandafter \@firstoftwo
 \else \expandafter \@secondoftwo
 \fi
}%
\providecommand \@ifx [1]{%
 \ifx #1\expandafter \@firstoftwo
 \else \expandafter \@secondoftwo
 \fi
}%
\providecommand \natexlab [1]{#1}%
\providecommand \enquote  [1]{``#1''}%
\providecommand \bibnamefont  [1]{#1}%
\providecommand \bibfnamefont [1]{#1}%
\providecommand \citenamefont [1]{#1}%
\providecommand \href@noop [0]{\@secondoftwo}%
\providecommand \href [0]{\begingroup \@sanitize@url \@href}%
\providecommand \@href[1]{\@@startlink{#1}\@@href}%
\providecommand \@@href[1]{\endgroup#1\@@endlink}%
\providecommand \@sanitize@url [0]{\catcode `\\12\catcode `\$12\catcode `\&12\catcode `\#12\catcode `\^12\catcode `\_12\catcode `\%12\relax}%
\providecommand \@@startlink[1]{}%
\providecommand \@@endlink[0]{}%
\providecommand \url  [0]{\begingroup\@sanitize@url \@url }%
\providecommand \@url [1]{\endgroup\@href {#1}{\urlprefix }}%
\providecommand \urlprefix  [0]{URL }%
\providecommand \Eprint [0]{\href }%
\providecommand \doibase [0]{https://doi.org/}%
\providecommand \selectlanguage [0]{\@gobble}%
\providecommand \bibinfo  [0]{\@secondoftwo}%
\providecommand \bibfield  [0]{\@secondoftwo}%
\providecommand \translation [1]{[#1]}%
\providecommand \BibitemOpen [0]{}%
\providecommand \bibitemStop [0]{}%
\providecommand \bibitemNoStop [0]{.\EOS\space}%
\providecommand \EOS [0]{\spacefactor3000\relax}%
\providecommand \BibitemShut  [1]{\csname bibitem#1\endcsname}%
\let\auto@bib@innerbib\@empty
\bibitem [{\citenamefont {Choi}\ \emph {et~al.}(2003)\citenamefont {Choi} \emph {et~al.}}]{Belle:2003nnu}%
  \BibitemOpen
  \bibfield  {author} {\bibinfo {author} {\bibfnamefont {S.~K.}\ \bibnamefont {Choi}} \emph {et~al.} (\bibinfo {collaboration} {Belle Collaboration}),\ }\href {https://doi.org/10.1103/PhysRevLett.91.262001} {\bibfield  {journal} {\bibinfo  {journal} {Phys. Rev. Lett.}\ }\textbf {\bibinfo {volume} {91}},\ \bibinfo {pages} {262001} (\bibinfo {year} {2003})}\BibitemShut {NoStop}%
\bibitem [{\citenamefont {Aubert}\ \emph {et~al.}(2005)\citenamefont {Aubert} \emph {et~al.}}]{BaBar:2005hhc}%
  \BibitemOpen
  \bibfield  {author} {\bibinfo {author} {\bibfnamefont {B.}~\bibnamefont {Aubert}} \emph {et~al.} (\bibinfo {collaboration} {BaBar Collaboration}),\ }\href {https://doi.org/10.1103/PhysRevLett.95.142001} {\bibfield  {journal} {\bibinfo  {journal} {Phys. Rev. Lett.}\ }\textbf {\bibinfo {volume} {95}},\ \bibinfo {pages} {142001} (\bibinfo {year} {2005})}\BibitemShut {NoStop}%
\bibitem [{\citenamefont {Ablikim}\ \emph {et~al.}(2013)\citenamefont {Ablikim} \emph {et~al.}}]{BESIII:2013ris}%
  \BibitemOpen
  \bibfield  {author} {\bibinfo {author} {\bibfnamefont {M.}~\bibnamefont {Ablikim}} \emph {et~al.} (\bibinfo {collaboration} {BESIII Collaboration}),\ }\href {https://doi.org/10.1103/PhysRevLett.110.252001} {\bibfield  {journal} {\bibinfo  {journal} {Phys. Rev. Lett.}\ }\textbf {\bibinfo {volume} {110}},\ \bibinfo {pages} {252001} (\bibinfo {year} {2013})}\BibitemShut {NoStop}%
\bibitem [{\citenamefont {Liu}\ \emph {et~al.}(2013)\citenamefont {Liu} \emph {et~al.}}]{Belle:2013yex}%
  \BibitemOpen
  \bibfield  {author} {\bibinfo {author} {\bibfnamefont {Z.~Q.}\ \bibnamefont {Liu}} \emph {et~al.} (\bibinfo {collaboration} {Belle Collaboration}),\ }\href {https://doi.org/10.1103/PhysRevLett.110.252002} {\bibfield  {journal} {\bibinfo  {journal} {Phys. Rev. Lett.}\ }\textbf {\bibinfo {volume} {110}},\ \bibinfo {pages} {252002} (\bibinfo {year} {2013})},\ \bibinfo {note} {[Erratum: Phys.Rev.Lett. 111, 019901 (2013)]}\BibitemShut {NoStop}%
\bibitem [{\citenamefont {Adams}\ \emph {et~al.}(1998)\citenamefont {Adams} \emph {et~al.}}]{E852:1998mbq}%
  \BibitemOpen
  \bibfield  {author} {\bibinfo {author} {\bibfnamefont {G.~S.}\ \bibnamefont {Adams}} \emph {et~al.} (\bibinfo {collaboration} {E852 Collaboration}),\ }\href {https://doi.org/10.1103/PhysRevLett.81.5760} {\bibfield  {journal} {\bibinfo  {journal} {Phys. Rev. Lett.}\ }\textbf {\bibinfo {volume} {81}},\ \bibinfo {pages} {5760} (\bibinfo {year} {1998})}\BibitemShut {NoStop}%
\bibitem [{\citenamefont {Aaij}\ \emph {et~al.}(2015)\citenamefont {Aaij} \emph {et~al.}}]{LHCb:2015yax}%
  \BibitemOpen
  \bibfield  {author} {\bibinfo {author} {\bibfnamefont {R.}~\bibnamefont {Aaij}} \emph {et~al.} (\bibinfo {collaboration} {LHCb Collaboration}),\ }\href {https://doi.org/10.1103/PhysRevLett.115.072001} {\bibfield  {journal} {\bibinfo  {journal} {Phys. Rev. Lett.}\ }\textbf {\bibinfo {volume} {115}},\ \bibinfo {pages} {072001} (\bibinfo {year} {2015})}\BibitemShut {NoStop}%
\bibitem [{\citenamefont {Guo}\ \emph {et~al.}(2018)\citenamefont {Guo} \emph {et~al.}}]{Guo:2017jvc}%
  \BibitemOpen
  \bibfield  {author} {\bibinfo {author} {\bibfnamefont {F.~K.}\ \bibnamefont {Guo}} \emph {et~al.},\ }\href {https://doi.org/10.1103/RevModPhys.90.015004} {\bibfield  {journal} {\bibinfo  {journal} {Rev. Mod. Phys.}\ }\textbf {\bibinfo {volume} {90}},\ \bibinfo {pages} {015004} (\bibinfo {year} {2018})}\BibitemShut {NoStop}%
\bibitem [{\citenamefont {Liu}\ \emph {et~al.}(2019)\citenamefont {Liu}, \citenamefont {Chen}, \citenamefont {Chen}, \citenamefont {Liu},\ and\ \citenamefont {Zhu}}]{Liu:2019zoy}%
  \BibitemOpen
  \bibfield  {author} {\bibinfo {author} {\bibfnamefont {Y.~R.}\ \bibnamefont {Liu}}, \bibinfo {author} {\bibfnamefont {H.~X.}\ \bibnamefont {Chen}}, \bibinfo {author} {\bibfnamefont {W.}~\bibnamefont {Chen}}, \bibinfo {author} {\bibfnamefont {X.}~\bibnamefont {Liu}},\ and\ \bibinfo {author} {\bibfnamefont {S.~L.}\ \bibnamefont {Zhu}},\ }\href {https://doi.org/10.1016/j.ppnp.2019.04.003} {\bibfield  {journal} {\bibinfo  {journal} {Prog. Part. Nucl. Phys.}\ }\textbf {\bibinfo {volume} {107}},\ \bibinfo {pages} {237} (\bibinfo {year} {2019})}\BibitemShut {NoStop}%
\bibitem [{\citenamefont {Brambilla}\ \emph {et~al.}(2020)\citenamefont {Brambilla} \emph {et~al.}}]{Brambilla:2019esw}%
  \BibitemOpen
  \bibfield  {author} {\bibinfo {author} {\bibfnamefont {N.}~\bibnamefont {Brambilla}} \emph {et~al.},\ }\href {https://doi.org/10.1016/j.physrep.2020.05.001} {\bibfield  {journal} {\bibinfo  {journal} {Phys. Rept.}\ }\textbf {\bibinfo {volume} {873}},\ \bibinfo {pages} {1} (\bibinfo {year} {2020})}\BibitemShut {NoStop}%
\bibitem [{\citenamefont {Klempt}\ and\ \citenamefont {Zaitsev}(2007)}]{Klempt:2007cp}%
  \BibitemOpen
  \bibfield  {author} {\bibinfo {author} {\bibfnamefont {E.}~\bibnamefont {Klempt}}\ and\ \bibinfo {author} {\bibfnamefont {A.}~\bibnamefont {Zaitsev}},\ }\href {https://doi.org/10.1016/j.physrep.2007.07.006} {\bibfield  {journal} {\bibinfo  {journal} {Phys. Rept.}\ }\textbf {\bibinfo {volume} {454}},\ \bibinfo {pages} {1} (\bibinfo {year} {2007})}\BibitemShut {NoStop}%
\bibitem [{\citenamefont {Jaffe}(2005)}]{Jaffe:2004ph}%
  \BibitemOpen
  \bibfield  {author} {\bibinfo {author} {\bibfnamefont {R.~L.}\ \bibnamefont {Jaffe}},\ }\href {https://doi.org/10.1016/j.physrep.2004.11.005} {\bibfield  {journal} {\bibinfo  {journal} {Phys. Rept.}\ }\textbf {\bibinfo {volume} {409}},\ \bibinfo {pages} {1} (\bibinfo {year} {2005})}\BibitemShut {NoStop}%
\bibitem [{\citenamefont {Locher}\ \emph {et~al.}(1986)\citenamefont {Locher}, \citenamefont {Sainio},\ and\ \citenamefont {Svarc}}]{Locher:1985nu}%
  \BibitemOpen
  \bibfield  {author} {\bibinfo {author} {\bibfnamefont {M.~P.}\ \bibnamefont {Locher}}, \bibinfo {author} {\bibfnamefont {M.~E.}\ \bibnamefont {Sainio}},\ and\ \bibinfo {author} {\bibfnamefont {A.}~\bibnamefont {Svarc}},\ }\href@noop {} {\bibfield  {journal} {\bibinfo  {journal} {Adv. Nucl. Phys.}\ }\textbf {\bibinfo {volume} {17}},\ \bibinfo {pages} {47} (\bibinfo {year} {1986})}\BibitemShut {NoStop}%
\bibitem [{\citenamefont {Abud}\ \emph {et~al.}(2010)\citenamefont {Abud}, \citenamefont {Buccella},\ and\ \citenamefont {Tramontano}}]{Abud:2009rk}%
  \BibitemOpen
  \bibfield  {author} {\bibinfo {author} {\bibfnamefont {M.}~\bibnamefont {Abud}}, \bibinfo {author} {\bibfnamefont {F.}~\bibnamefont {Buccella}},\ and\ \bibinfo {author} {\bibfnamefont {F.}~\bibnamefont {Tramontano}},\ }\href {https://doi.org/10.1103/PhysRevD.81.074018} {\bibfield  {journal} {\bibinfo  {journal} {Phys. Rev. D}\ }\textbf {\bibinfo {volume} {81}},\ \bibinfo {pages} {074018} (\bibinfo {year} {2010})}\BibitemShut {NoStop}%
\bibitem [{\citenamefont {Bashkanov}\ \emph {et~al.}(2013)\citenamefont {Bashkanov}, \citenamefont {Brodsky},\ and\ \citenamefont {Clement}}]{Bashkanov:2013cla}%
  \BibitemOpen
  \bibfield  {author} {\bibinfo {author} {\bibfnamefont {M.}~\bibnamefont {Bashkanov}}, \bibinfo {author} {\bibfnamefont {S.~J.}\ \bibnamefont {Brodsky}},\ and\ \bibinfo {author} {\bibfnamefont {H.}~\bibnamefont {Clement}},\ }\href {https://doi.org/10.1016/j.physletb.2013.10.059} {\bibfield  {journal} {\bibinfo  {journal} {Phys. Lett. B}\ }\textbf {\bibinfo {volume} {727}},\ \bibinfo {pages} {438} (\bibinfo {year} {2013})}\BibitemShut {NoStop}%
\bibitem [{\citenamefont {Clement}(2017)}]{Clement:2016vnl}%
  \BibitemOpen
  \bibfield  {author} {\bibinfo {author} {\bibfnamefont {H.}~\bibnamefont {Clement}},\ }\href {https://doi.org/10.1016/j.ppnp.2016.12.004} {\bibfield  {journal} {\bibinfo  {journal} {Prog. Part. Nucl. Phys.}\ }\textbf {\bibinfo {volume} {93}},\ \bibinfo {pages} {195} (\bibinfo {year} {2017})}\BibitemShut {NoStop}%
\bibitem [{\citenamefont {Gal}(2016)}]{Gal:2015rev}%
  \BibitemOpen
  \bibfield  {author} {\bibinfo {author} {\bibfnamefont {A.}~\bibnamefont {Gal}},\ }\href {https://doi.org/10.5506/APhysPolB.47.471} {\bibfield  {journal} {\bibinfo  {journal} {Acta Phys. Polon. B}\ }\textbf {\bibinfo {volume} {47}},\ \bibinfo {pages} {471} (\bibinfo {year} {2016})}\BibitemShut {NoStop}%
\bibitem [{\citenamefont {Dong}\ \emph {et~al.}(2023)\citenamefont {Dong}, \citenamefont {Shen},\ and\ \citenamefont {Zhang}}]{Dong:2023xdi}%
  \BibitemOpen
  \bibfield  {author} {\bibinfo {author} {\bibfnamefont {Y.}~\bibnamefont {Dong}}, \bibinfo {author} {\bibfnamefont {P.}~\bibnamefont {Shen}},\ and\ \bibinfo {author} {\bibfnamefont {Z.}~\bibnamefont {Zhang}},\ }\href {https://doi.org/10.1016/j.ppnp.2023.104045} {\bibfield  {journal} {\bibinfo  {journal} {Prog. Part. Nucl. Phys.}\ }\textbf {\bibinfo {volume} {131}},\ \bibinfo {pages} {104045} (\bibinfo {year} {2023})}\BibitemShut {NoStop}%
\bibitem [{\citenamefont {Bashkanov}\ \emph {et~al.}(2009)\citenamefont {Bashkanov} \emph {et~al.}}]{Bashkanov:2008ih}%
  \BibitemOpen
  \bibfield  {author} {\bibinfo {author} {\bibfnamefont {M.}~\bibnamefont {Bashkanov}} \emph {et~al.} (\bibinfo {collaboration} {CELSIUS/WASA Collaboration}),\ }\href {https://doi.org/10.1103/PhysRevLett.102.052301} {\bibfield  {journal} {\bibinfo  {journal} {Phys. Rev. Lett.}\ }\textbf {\bibinfo {volume} {102}},\ \bibinfo {pages} {052301} (\bibinfo {year} {2009})}\BibitemShut {NoStop}%
\bibitem [{\citenamefont {Adlarson}\ \emph {et~al.}(2011)\citenamefont {Adlarson} \emph {et~al.}}]{WASA-at-COSY:2011bjg}%
  \BibitemOpen
  \bibfield  {author} {\bibinfo {author} {\bibfnamefont {P.}~\bibnamefont {Adlarson}} \emph {et~al.} (\bibinfo {collaboration} {WASA-at-COSY Collaboration}),\ }\href {https://doi.org/10.1103/PhysRevLett.106.242302} {\bibfield  {journal} {\bibinfo  {journal} {Phys. Rev. Lett.}\ }\textbf {\bibinfo {volume} {106}},\ \bibinfo {pages} {242302} (\bibinfo {year} {2011})}\BibitemShut {NoStop}%
\bibitem [{\citenamefont {Kren}\ \emph {et~al.}(2010)\citenamefont {Kren} \emph {et~al.}}]{CELSIUSWASA:2009dos}%
  \BibitemOpen
  \bibfield  {author} {\bibinfo {author} {\bibfnamefont {F.}~\bibnamefont {Kren}} \emph {et~al.} (\bibinfo {collaboration} {CELSIUS/WASA Collaboration}),\ }\href {https://doi.org/10.1016/j.physletb.2009.12.061} {\bibfield  {journal} {\bibinfo  {journal} {Phys. Lett. B}\ }\textbf {\bibinfo {volume} {684}},\ \bibinfo {pages} {110} (\bibinfo {year} {2010})},\ \bibinfo {note} {[Erratum: Phys.Lett.B 702, 312--313 (2011)]}\BibitemShut {NoStop}%
\bibitem [{\citenamefont {Adlarson}\ \emph {et~al.}(2013)\citenamefont {Adlarson} \emph {et~al.}}]{WASA-at-COSY:2013fzt}%
  \BibitemOpen
  \bibfield  {author} {\bibinfo {author} {\bibfnamefont {P.}~\bibnamefont {Adlarson}} \emph {et~al.} (\bibinfo {collaboration} {WASA-at-COSY Collaboration}),\ }\href {https://doi.org/10.1103/PhysRevC.88.055208} {\bibfield  {journal} {\bibinfo  {journal} {Phys. Rev. C}\ }\textbf {\bibinfo {volume} {88}},\ \bibinfo {pages} {055208} (\bibinfo {year} {2013})}\BibitemShut {NoStop}%
\bibitem [{\citenamefont {Adlarson}\ \emph {et~al.}(2015)\citenamefont {Adlarson} \emph {et~al.}}]{WASA-at-COSY:2014qkg}%
  \BibitemOpen
  \bibfield  {author} {\bibinfo {author} {\bibfnamefont {P.}~\bibnamefont {Adlarson}} \emph {et~al.} (\bibinfo {collaboration} {WASA-at-COSY Collaboration}),\ }\href {https://doi.org/10.1016/j.physletb.2015.02.067} {\bibfield  {journal} {\bibinfo  {journal} {Phys. Lett. B}\ }\textbf {\bibinfo {volume} {743}},\ \bibinfo {pages} {325} (\bibinfo {year} {2015})}\BibitemShut {NoStop}%
\bibitem [{\citenamefont {Huang}\ \emph {et~al.}(2014)\citenamefont {Huang}, \citenamefont {Ping},\ and\ \citenamefont {Wang}}]{Huang:2013nba}%
  \BibitemOpen
  \bibfield  {author} {\bibinfo {author} {\bibfnamefont {H.~X.}\ \bibnamefont {Huang}}, \bibinfo {author} {\bibfnamefont {J.~L.}\ \bibnamefont {Ping}},\ and\ \bibinfo {author} {\bibfnamefont {F.}~\bibnamefont {Wang}},\ }\href {https://doi.org/10.1103/PhysRevC.89.034001} {\bibfield  {journal} {\bibinfo  {journal} {Phys. Rev. C}\ }\textbf {\bibinfo {volume} {89}},\ \bibinfo {pages} {034001} (\bibinfo {year} {2014})}\BibitemShut {NoStop}%
\bibitem [{\citenamefont {Kim}\ \emph {et~al.}(2020)\citenamefont {Kim}, \citenamefont {Kim},\ and\ \citenamefont {Oka}}]{Kim:2020rwn}%
  \BibitemOpen
  \bibfield  {author} {\bibinfo {author} {\bibfnamefont {H.}~\bibnamefont {Kim}}, \bibinfo {author} {\bibfnamefont {K.~S.}\ \bibnamefont {Kim}},\ and\ \bibinfo {author} {\bibfnamefont {M.}~\bibnamefont {Oka}},\ }\href {https://doi.org/10.1103/PhysRevD.102.074023} {\bibfield  {journal} {\bibinfo  {journal} {Phys. Rev. D}\ }\textbf {\bibinfo {volume} {102}},\ \bibinfo {pages} {074023} (\bibinfo {year} {2020})}\BibitemShut {NoStop}%
\bibitem [{\citenamefont {Bashkanov}\ \emph {et~al.}(2020)\citenamefont {Bashkanov} \emph {et~al.}}]{A2:2019arr}%
  \BibitemOpen
  \bibfield  {author} {\bibinfo {author} {\bibfnamefont {M.}~\bibnamefont {Bashkanov}} \emph {et~al.} (\bibinfo {collaboration} {A2 Collaboration}),\ }\href {https://doi.org/10.1103/PhysRevLett.124.132001} {\bibfield  {journal} {\bibinfo  {journal} {Phys. Rev. Lett.}\ }\textbf {\bibinfo {volume} {124}},\ \bibinfo {pages} {132001} (\bibinfo {year} {2020})}\BibitemShut {NoStop}%
\bibitem [{\citenamefont {Albrecht}\ \emph {et~al.}(1990)\citenamefont {Albrecht} \emph {et~al.}}]{ARGUS:1989sto}%
  \BibitemOpen
  \bibfield  {author} {\bibinfo {author} {\bibfnamefont {H.}~\bibnamefont {Albrecht}} \emph {et~al.} (\bibinfo {collaboration} {ARGUS Collaboration}),\ }\href {https://doi.org/10.1016/0370-2693(90)90602-3} {\bibfield  {journal} {\bibinfo  {journal} {Phys. Lett. B}\ }\textbf {\bibinfo {volume} {236}},\ \bibinfo {pages} {102} (\bibinfo {year} {1990})}\BibitemShut {NoStop}%
\bibitem [{\citenamefont {Asner}\ \emph {et~al.}(2007)\citenamefont {Asner} \emph {et~al.}}]{CLEO:2006zjy}%
  \BibitemOpen
  \bibfield  {author} {\bibinfo {author} {\bibfnamefont {D.~M.}\ \bibnamefont {Asner}} \emph {et~al.} (\bibinfo {collaboration} {CLEO Collaboration}),\ }\href {https://doi.org/10.1103/PhysRevD.75.012009} {\bibfield  {journal} {\bibinfo  {journal} {Phys. Rev. D}\ }\textbf {\bibinfo {volume} {75}},\ \bibinfo {pages} {012009} (\bibinfo {year} {2007})}\BibitemShut {NoStop}%
\bibitem [{\citenamefont {Lees}\ \emph {et~al.}(2014)\citenamefont {Lees} \emph {et~al.}}]{BaBar:2014ssg}%
  \BibitemOpen
  \bibfield  {author} {\bibinfo {author} {\bibfnamefont {J.~P.}\ \bibnamefont {Lees}} \emph {et~al.} (\bibinfo {collaboration} {BaBar Collaboration}),\ }\href {https://doi.org/10.1103/PhysRevD.89.111102} {\bibfield  {journal} {\bibinfo  {journal} {Phys. Rev. D}\ }\textbf {\bibinfo {volume} {89}},\ \bibinfo {pages} {111102} (\bibinfo {year} {2014})}\BibitemShut {NoStop}%
\bibitem [{\citenamefont {Schael}\ \emph {et~al.}(2006)\citenamefont {Schael} \emph {et~al.}}]{ALEPH:2006qoi}%
  \BibitemOpen
  \bibfield  {author} {\bibinfo {author} {\bibfnamefont {S.}~\bibnamefont {Schael}} \emph {et~al.} (\bibinfo {collaboration} {ALEPH Collaboration}),\ }\href {https://doi.org/10.1016/j.physletb.2006.06.043} {\bibfield  {journal} {\bibinfo  {journal} {Phys. Lett. B}\ }\textbf {\bibinfo {volume} {639}},\ \bibinfo {pages} {192} (\bibinfo {year} {2006})}\BibitemShut {NoStop}%
\bibitem [{\citenamefont {Ablikim}\ \emph {et~al.}(2010)\citenamefont {Ablikim} \emph {et~al.}}]{BESIII:2009fln}%
  \BibitemOpen
  \bibfield  {author} {\bibinfo {author} {\bibfnamefont {M.}~\bibnamefont {Ablikim}} \emph {et~al.} (\bibinfo {collaboration} {BESIII Collaboration}),\ }\href {https://doi.org/10.1016/j.nima.2009.12.050} {\bibfield  {journal} {\bibinfo  {journal} {Nucl. Instrum. Meth. A}\ }\textbf {\bibinfo {volume} {614}},\ \bibinfo {pages} {345} (\bibinfo {year} {2010})}\BibitemShut {NoStop}%
\bibitem [{\citenamefont {Ablikim}\ \emph {et~al.}(2021{\natexlab{a}})\citenamefont {Ablikim} \emph {et~al.}}]{BESIII:2020svk}%
  \BibitemOpen
  \bibfield  {author} {\bibinfo {author} {\bibfnamefont {M.}~\bibnamefont {Ablikim}} \emph {et~al.} (\bibinfo {collaboration} {BESIII Collaboration}),\ }\href {https://doi.org/10.1103/PhysRevD.103.052003} {\bibfield  {journal} {\bibinfo  {journal} {Phys. Rev. D}\ }\textbf {\bibinfo {volume} {103}},\ \bibinfo {pages} {052003} (\bibinfo {year} {2021}{\natexlab{a}})}\BibitemShut {NoStop}%
\bibitem [{\citenamefont {Ablikim}\ \emph {et~al.}(2023)\citenamefont {Ablikim} \emph {et~al.}}]{BESIII:pppnpi}%
  \BibitemOpen
  \bibfield  {author} {\bibinfo {author} {\bibfnamefont {M.}~\bibnamefont {Ablikim}} \emph {et~al.} (\bibinfo {collaboration} {BESIII Collaboration}),\ }\href {https://doi.org/10.1088/1674-1137/acb6eb} {\bibfield  {journal} {\bibinfo  {journal} {Chin. Phys. C}\ }\textbf {\bibinfo {volume} {47}},\ \bibinfo {pages} {043001} (\bibinfo {year} {2023})}\BibitemShut {NoStop}%
\bibitem [{\citenamefont {Yu}\ \emph {et~al.}(2016)\citenamefont {Yu} \emph {et~al.}}]{Yu:IPAC2016-TUYA01}%
  \BibitemOpen
  \bibfield  {author} {\bibinfo {author} {\bibfnamefont {C.~h.}\ \bibnamefont {Yu}} \emph {et~al.},\ }in\ \href {https://doi.org/10.18429/JACoW-IPAC2016-TUYA01} {\emph {\bibinfo {booktitle} {{7th International Particle Accelerator Conference}}}}\ (\bibinfo {year} {2016})\ p.\ \bibinfo {pages} {TUYA01}\BibitemShut {NoStop}%
\bibitem [{\citenamefont {Ablikim}\ \emph {et~al.}(2020)\citenamefont {Ablikim} \emph {et~al.}}]{BESIII:2020nme}%
  \BibitemOpen
  \bibfield  {author} {\bibinfo {author} {\bibfnamefont {M.}~\bibnamefont {Ablikim}} \emph {et~al.} (\bibinfo {collaboration} {BESIII Collaboration}),\ }\href {https://doi.org/10.1088/1674-1137/44/4/040001} {\bibfield  {journal} {\bibinfo  {journal} {Chin. Phys. C}\ }\textbf {\bibinfo {volume} {44}},\ \bibinfo {pages} {040001} (\bibinfo {year} {2020})}\BibitemShut {NoStop}%
\bibitem [{\citenamefont {Huang}\ \emph {et~al.}(2022)\citenamefont {Huang}, \citenamefont {Li}, \citenamefont {Qian}, \citenamefont {Zhu}, \citenamefont {Li}, \citenamefont {Zhang}, \citenamefont {Sun},\ and\ \citenamefont {You}}]{Huang:2022wuo}%
  \BibitemOpen
  \bibfield  {author} {\bibinfo {author} {\bibfnamefont {K.~X.}\ \bibnamefont {Huang}}, \bibinfo {author} {\bibfnamefont {Z.~J.}\ \bibnamefont {Li}}, \bibinfo {author} {\bibfnamefont {Z.}~\bibnamefont {Qian}}, \bibinfo {author} {\bibfnamefont {J.}~\bibnamefont {Zhu}}, \bibinfo {author} {\bibfnamefont {H.~Y.}\ \bibnamefont {Li}}, \bibinfo {author} {\bibfnamefont {Y.~M.}\ \bibnamefont {Zhang}}, \bibinfo {author} {\bibfnamefont {S.~S.}\ \bibnamefont {Sun}},\ and\ \bibinfo {author} {\bibfnamefont {Z.~Y.}\ \bibnamefont {You}},\ }\href {https://doi.org/10.1007/s41365-022-01133-8} {\bibfield  {journal} {\bibinfo  {journal} {Nucl. Sci. Tech.}\ }\textbf {\bibinfo {volume} {33}},\ \bibinfo {pages} {142} (\bibinfo {year} {2022})}\BibitemShut {NoStop}%
\bibitem [{\citenamefont {Li}\ \emph {et~al.}(2017)\citenamefont {Li} \emph {et~al.}}]{etof1}%
  \BibitemOpen
  \bibfield  {author} {\bibinfo {author} {\bibfnamefont {X.}~\bibnamefont {Li}} \emph {et~al.},\ }\href {https://doi.org/10.1007/s41605-017-0014-2} {\bibfield  {journal} {\bibinfo  {journal} {Radiat. Detect. Technol. Methods}\ }\textbf {\bibinfo {volume} {1}} (\bibinfo {year} {2017})}\BibitemShut {NoStop}%
\bibitem [{\citenamefont {Guo}\ \emph {et~al.}(2017)\citenamefont {Guo} \emph {et~al.}}]{etof2}%
  \BibitemOpen
  \bibfield  {author} {\bibinfo {author} {\bibfnamefont {Y.~X.}\ \bibnamefont {Guo}} \emph {et~al.},\ }\href {https://doi.org/10.1007/s41605-017-0012-4} {\bibfield  {journal} {\bibinfo  {journal} {Radiat. Detect. Technol. Methods}\ }\textbf {\bibinfo {volume} {1}},\ \bibinfo {pages} {1} (\bibinfo {year} {2017})}\BibitemShut {NoStop}%
\bibitem [{\citenamefont {Cao}\ \emph {et~al.}(2020)\citenamefont {Cao} \emph {et~al.}}]{etof3}%
  \BibitemOpen
  \bibfield  {author} {\bibinfo {author} {\bibfnamefont {P.}~\bibnamefont {Cao}} \emph {et~al.},\ }\href {https://doi.org/10.1016/j.nima.2019.163053} {\bibfield  {journal} {\bibinfo  {journal} {Nucl. Instrum. Meth. A}\ }\textbf {\bibinfo {volume} {953}},\ \bibinfo {pages} {163053} (\bibinfo {year} {2020})}\BibitemShut {NoStop}%
\bibitem [{\citenamefont {Ablikim}\ \emph {et~al.}(2016)\citenamefont {Ablikim} \emph {et~al.}}]{BESIII:2015zbz}%
  \BibitemOpen
  \bibfield  {author} {\bibinfo {author} {\bibfnamefont {M.}~\bibnamefont {Ablikim}} \emph {et~al.} (\bibinfo {collaboration} {BESIII Collaboration}),\ }\href {https://doi.org/10.1088/1674-1137/40/6/063001} {\bibfield  {journal} {\bibinfo  {journal} {Chin. Phys. C}\ }\textbf {\bibinfo {volume} {40}},\ \bibinfo {pages} {063001} (\bibinfo {year} {2016})}\BibitemShut {NoStop}%
\bibitem [{\citenamefont {Ablikim}\ \emph {et~al.}(2021{\natexlab{b}})\citenamefont {Ablikim} \emph {et~al.}}]{BESIII:2020eyu}%
  \BibitemOpen
  \bibfield  {author} {\bibinfo {author} {\bibfnamefont {M.}~\bibnamefont {Ablikim}} \emph {et~al.} (\bibinfo {collaboration} {BESIII Collaboration}),\ }\href {https://doi.org/10.1088/1674-1137/ac1575} {\bibfield  {journal} {\bibinfo  {journal} {Chin. Phys. C}\ }\textbf {\bibinfo {volume} {45}},\ \bibinfo {pages} {103001} (\bibinfo {year} {2021}{\natexlab{b}})}\BibitemShut {NoStop}%
\bibitem [{\citenamefont {Ablikim}\ \emph {et~al.}(2022{\natexlab{a}})\citenamefont {Ablikim} \emph {et~al.}}]{BESIII:2022ulv}%
  \BibitemOpen
  \bibfield  {author} {\bibinfo {author} {\bibfnamefont {M.}~\bibnamefont {Ablikim}} \emph {et~al.} (\bibinfo {collaboration} {BESIII}),\ }\href {https://doi.org/10.1088/1674-1137/ac84cc} {\bibfield  {journal} {\bibinfo  {journal} {Chin. Phys. C}\ }\textbf {\bibinfo {volume} {46}},\ \bibinfo {pages} {113003} (\bibinfo {year} {2022}{\natexlab{a}})}\BibitemShut {NoStop}%
\bibitem [{\citenamefont {Ablikim}\ \emph {et~al.}(2022{\natexlab{b}})\citenamefont {Ablikim} \emph {et~al.}}]{BESIII:2022dxl}%
  \BibitemOpen
  \bibfield  {author} {\bibinfo {author} {\bibfnamefont {M.}~\bibnamefont {Ablikim}} \emph {et~al.} (\bibinfo {collaboration} {BESIII}),\ }\href {https://doi.org/10.1088/1674-1137/ac80b4} {\bibfield  {journal} {\bibinfo  {journal} {Chin. Phys. C}\ }\textbf {\bibinfo {volume} {46}},\ \bibinfo {pages} {113002} (\bibinfo {year} {2022}{\natexlab{b}})}\BibitemShut {NoStop}%
\bibitem [{\citenamefont {Ablikim}\ \emph {et~al.}(2015)\citenamefont {Ablikim} \emph {et~al.}}]{BESIII:2015qfd}%
  \BibitemOpen
  \bibfield  {author} {\bibinfo {author} {\bibfnamefont {M.}~\bibnamefont {Ablikim}} \emph {et~al.} (\bibinfo {collaboration} {BESIII Collaboration}),\ }\href {https://doi.org/10.1088/1674-1137/39/9/093001} {\bibfield  {journal} {\bibinfo  {journal} {Chin. Phys. C}\ }\textbf {\bibinfo {volume} {39}},\ \bibinfo {pages} {093001} (\bibinfo {year} {2015})}\BibitemShut {NoStop}%
\bibitem [{\citenamefont {Agostinelli}\ \emph {et~al.}(2003)\citenamefont {Agostinelli} \emph {et~al.}}]{GEANT4:2002zbu}%
  \BibitemOpen
  \bibfield  {author} {\bibinfo {author} {\bibfnamefont {S.}~\bibnamefont {Agostinelli}} \emph {et~al.} (\bibinfo {collaboration} {GEANT4 Collaboration}),\ }\href {https://doi.org/10.1016/S0168-9002(03)01368-8} {\bibfield  {journal} {\bibinfo  {journal} {Nucl. Instrum. Meth. A}\ }\textbf {\bibinfo {volume} {506}},\ \bibinfo {pages} {250} (\bibinfo {year} {2003})}\BibitemShut {NoStop}%
\bibitem [{\citenamefont {Jadach}\ \emph {et~al.}(2001)\citenamefont {Jadach}, \citenamefont {Ward},\ and\ \citenamefont {Was}}]{Jadach:2000ir}%
  \BibitemOpen
  \bibfield  {author} {\bibinfo {author} {\bibfnamefont {S.}~\bibnamefont {Jadach}}, \bibinfo {author} {\bibfnamefont {B.~F.~L.}\ \bibnamefont {Ward}},\ and\ \bibinfo {author} {\bibfnamefont {Z.}~\bibnamefont {Was}},\ }\href {https://doi.org/10.1103/PhysRevD.63.113009} {\bibfield  {journal} {\bibinfo  {journal} {Phys. Rev. D}\ }\textbf {\bibinfo {volume} {63}},\ \bibinfo {pages} {113009} (\bibinfo {year} {2001})}\BibitemShut {NoStop}%
\bibitem [{\citenamefont {Jadach}\ \emph {et~al.}(2000)\citenamefont {Jadach}, \citenamefont {Ward},\ and\ \citenamefont {Was}}]{Jadach:1999vf}%
  \BibitemOpen
  \bibfield  {author} {\bibinfo {author} {\bibfnamefont {S.}~\bibnamefont {Jadach}}, \bibinfo {author} {\bibfnamefont {B.~F.~L.}\ \bibnamefont {Ward}},\ and\ \bibinfo {author} {\bibfnamefont {Z.}~\bibnamefont {Was}},\ }\href {https://doi.org/10.1016/S0010-4655(00)00048-5} {\bibfield  {journal} {\bibinfo  {journal} {Comput. Phys. Commun.}\ }\textbf {\bibinfo {volume} {130}},\ \bibinfo {pages} {260} (\bibinfo {year} {2000})}\BibitemShut {NoStop}%
\bibitem [{\citenamefont {Lange}(2001)}]{Lange:2001uf}%
  \BibitemOpen
  \bibfield  {author} {\bibinfo {author} {\bibfnamefont {D.~J.}\ \bibnamefont {Lange}},\ }\href {https://doi.org/10.1016/S0168-9002(01)00089-4} {\bibfield  {journal} {\bibinfo  {journal} {Nucl. Instrum. Meth. A}\ }\textbf {\bibinfo {volume} {462}},\ \bibinfo {pages} {152} (\bibinfo {year} {2001})}\BibitemShut {NoStop}%
\bibitem [{\citenamefont {Ping}(2008)}]{Ping:2008zz}%
  \BibitemOpen
  \bibfield  {author} {\bibinfo {author} {\bibfnamefont {R.~G.}\ \bibnamefont {Ping}},\ }\href {https://doi.org/10.1088/1674-1137/32/8/001} {\bibfield  {journal} {\bibinfo  {journal} {Chin. Phys. C}\ }\textbf {\bibinfo {volume} {32}},\ \bibinfo {pages} {599} (\bibinfo {year} {2008})}\BibitemShut {NoStop}%
\bibitem [{\citenamefont {Workman}\ \emph {et~al.}(2022)\citenamefont {Workman} \emph {et~al.}}]{pdg}%
  \BibitemOpen
  \bibfield  {author} {\bibinfo {author} {\bibfnamefont {R.~L.}\ \bibnamefont {Workman}} \emph {et~al.} (\bibinfo {collaboration} {Particle Data Group}),\ }\href {https://doi.org/10.1093/ptep/ptac097} {\bibfield  {journal} {\bibinfo  {journal} {PTEP}\ }\textbf {\bibinfo {volume} {2022}},\ \bibinfo {pages} {083C01} (\bibinfo {year} {2022})}\BibitemShut {NoStop}%
\bibitem [{\citenamefont {Chen}\ \emph {et~al.}(2000)\citenamefont {Chen}, \citenamefont {Huang}, \citenamefont {Qi}, \citenamefont {Zhang},\ and\ \citenamefont {Zhu}}]{Chen:2000tv}%
  \BibitemOpen
  \bibfield  {author} {\bibinfo {author} {\bibfnamefont {J.~C.}\ \bibnamefont {Chen}}, \bibinfo {author} {\bibfnamefont {G.~S.}\ \bibnamefont {Huang}}, \bibinfo {author} {\bibfnamefont {X.~R.}\ \bibnamefont {Qi}}, \bibinfo {author} {\bibfnamefont {D.~H.}\ \bibnamefont {Zhang}},\ and\ \bibinfo {author} {\bibfnamefont {Y.~S.}\ \bibnamefont {Zhu}},\ }\href {https://doi.org/10.1103/PhysRevD.62.034003} {\bibfield  {journal} {\bibinfo  {journal} {Phys. Rev. D}\ }\textbf {\bibinfo {volume} {62}},\ \bibinfo {pages} {034003} (\bibinfo {year} {2000})}\BibitemShut {NoStop}%
\bibitem [{\citenamefont {Yang}\ \emph {et~al.}(2014)\citenamefont {Yang}, \citenamefont {Ping},\ and\ \citenamefont {Chen}}]{Yang:2014vra}%
  \BibitemOpen
  \bibfield  {author} {\bibinfo {author} {\bibfnamefont {R.~L.}\ \bibnamefont {Yang}}, \bibinfo {author} {\bibfnamefont {R.~G.}\ \bibnamefont {Ping}},\ and\ \bibinfo {author} {\bibfnamefont {H.}~\bibnamefont {Chen}},\ }\href {https://doi.org/10.1088/0256-307X/31/6/061301} {\bibfield  {journal} {\bibinfo  {journal} {Chin. Phys. Lett.}\ }\textbf {\bibinfo {volume} {31}},\ \bibinfo {pages} {061301} (\bibinfo {year} {2014})}\BibitemShut {NoStop}%
\bibitem [{\citenamefont {Richter~Was}(1993)}]{Richter-Was:1992hxq}%
  \BibitemOpen
  \bibfield  {author} {\bibinfo {author} {\bibfnamefont {E.}~\bibnamefont {Richter~Was}},\ }\href {https://doi.org/10.1016/0370-2693(93)90062-M} {\bibfield  {journal} {\bibinfo  {journal} {Phys. Lett. B}\ }\textbf {\bibinfo {volume} {303}},\ \bibinfo {pages} {163} (\bibinfo {year} {1993})}\BibitemShut {NoStop}%
\bibitem [{\citenamefont {Rolke}\ \emph {et~al.}(2005)\citenamefont {Rolke}, \citenamefont {Lopez},\ and\ \citenamefont {Conrad}}]{Rolke:2004mj}%
  \BibitemOpen
  \bibfield  {author} {\bibinfo {author} {\bibfnamefont {W.~A.}\ \bibnamefont {Rolke}}, \bibinfo {author} {\bibfnamefont {A.~M.}\ \bibnamefont {Lopez}},\ and\ \bibinfo {author} {\bibfnamefont {J.}~\bibnamefont {Conrad}},\ }\href {https://doi.org/10.1016/j.nima.2005.05.068} {\bibfield  {journal} {\bibinfo  {journal} {Nucl. Instrum. Meth. A}\ }\textbf {\bibinfo {volume} {551}},\ \bibinfo {pages} {493} (\bibinfo {year} {2005})}\BibitemShut {NoStop}%
\bibitem [{\citenamefont {Lundberg}\ \emph {et~al.}(2010)\citenamefont {Lundberg}, \citenamefont {Conrad}, \citenamefont {Rolke},\ and\ \citenamefont {Lopez}}]{Lundberg:2009iu}%
  \BibitemOpen
  \bibfield  {author} {\bibinfo {author} {\bibfnamefont {J.}~\bibnamefont {Lundberg}}, \bibinfo {author} {\bibfnamefont {J.}~\bibnamefont {Conrad}}, \bibinfo {author} {\bibfnamefont {W.}~\bibnamefont {Rolke}},\ and\ \bibinfo {author} {\bibfnamefont {A.}~\bibnamefont {Lopez}},\ }\href {https://doi.org/10.1016/j.cpc.2009.11.001} {\bibfield  {journal} {\bibinfo  {journal} {Comput. Phys. Commun.}\ }\textbf {\bibinfo {volume} {181}},\ \bibinfo {pages} {683} (\bibinfo {year} {2010})}\BibitemShut {NoStop}%
\bibitem [{\citenamefont {Brun}\ and\ \citenamefont {Rademakers}(1997)}]{BRUN199781}%
  \BibitemOpen
  \bibfield  {author} {\bibinfo {author} {\bibfnamefont {R.}~\bibnamefont {Brun}}\ and\ \bibinfo {author} {\bibfnamefont {F.}~\bibnamefont {Rademakers}},\ }\href {https://doi.org/https://doi.org/10.1016/S0168-9002(97)00048-X} {\bibfield  {journal} {\bibinfo  {journal} {Nucl. Instrum. Meth. A}\ }\textbf {\bibinfo {volume} {389}},\ \bibinfo {pages} {81} (\bibinfo {year} {1997})}\BibitemShut {NoStop}%
\bibitem [{\citenamefont {Ablikim}\ \emph {et~al.}(2021{\natexlab{c}})\citenamefont {Ablikim} \emph {et~al.}}]{BESIII:2020lkm}%
  \BibitemOpen
  \bibfield  {author} {\bibinfo {author} {\bibfnamefont {M.}~\bibnamefont {Ablikim}} \emph {et~al.} (\bibinfo {collaboration} {BESIII Collaboration}),\ }\href {https://doi.org/10.1103/PhysRevLett.126.092002} {\bibfield  {journal} {\bibinfo  {journal} {Phys. Rev. Lett.}\ }\textbf {\bibinfo {volume} {126}},\ \bibinfo {pages} {092002} (\bibinfo {year} {2021}{\natexlab{c}})}\BibitemShut {NoStop}%
\bibitem [{\citenamefont {Jegerlehner}(2011)}]{vacuum}%
  \BibitemOpen
  \bibfield  {author} {\bibinfo {author} {\bibfnamefont {F.}~\bibnamefont {Jegerlehner}},\ }\href {https://doi.org/10.1393/ncc/i2011-11011-0} {\bibfield  {journal} {\bibinfo  {journal} {Nuovo Cimento C}\ }\textbf {\bibinfo {volume} {034S1}},\ \bibinfo {pages} {31} (\bibinfo {year} {2011})}\BibitemShut {NoStop}%
\end{thebibliography}%

\end{document}